\documentclass[useAMS,usenatbib]{mnras}

\usepackage{amsfonts}
\usepackage{amsmath}
\usepackage{epsfig}
\usepackage{amssymb}
\usepackage{gensymb}
\usepackage{amsfonts}
\usepackage{amsmath}

\usepackage[T1]{fontenc}
\usepackage{ae,aecompl}
\usepackage{subfigure}

\usepackage{graphicx}	
\usepackage{amsmath}	
\usepackage{amssymb}	

\title[Gaia DR2 lensed quasars]{Gravitationally lensed quasars in \textit{Gaia}: III. 22 new lensed quasars from \textit{Gaia} Data Release 2}

\author[C. A. Lemon et al.]{
Cameron A. Lemon,$^{1, 2}$\thanks{E-mail: clemon@ast.cam.ac.uk}
Matthew W. Auger$^{1, 2}$,
Richard G. McMahon$^{1, 2}$
\\
$^{1}$Institute of Astronomy, University of Cambridge, Madingley Road, Cambridge CB3 0HA, UK\\
$^{2}$Kavli  Institute  for  Cosmology,  University  of  Cambridge,  Madingley Road, Cambridge CB3 0HA, UK
}

\date{Accepted XXX. Received YYY; in original form ZZZ}

\pubyear{2015}

\begin{document}
\label{firstpage}
\pagerange{\pageref{firstpage}--\pageref{lastpage}}
\maketitle
\newcommand{\MINUS}{\kern 0.102em --\kern 0.102em }
\newcommand{\Gaia}{\textit{Gaia} }
\begin{abstract}
We report the discovery and spectroscopic confirmation of 22 new gravitationally lensed quasars found using \Gaia data release 2. The selection was made using several techniques: multiple \Gaia detections around objects in quasar candidate catalogues, modelling of unWISE coadd pixels using \Gaia astrometry, and \Gaia detections offset from photometric and spectroscopic galaxies. Spectra of 33 candidates were obtained with the William Herschel Telescope, 22 of which are lensed quasars, 2 highly probably lensed quasars, 5 nearly identical quasar pairs, 1 inconclusive system, and 3 contaminants. Of the 3 confirmed quadruply imaged systems, J2145+6345 is a 2.1 arcsecond separation quad with 4 bright images ($G$=16.86, 17.26, 18.34, 18.56), making it ideal for time delay monitoring. Analysing this new sample alongside known lenses in the Pan-STARRS footprint, and comparing to expected numbers of lenses, we show that, as expected, we are biased towards systems with bright lensing galaxies and low source redshifts. We discuss possible techniques to remove this bias from future searches. A |b|>20 complete sample of lensed quasars detected by \Gaia and with image separations above 1 arcsecond will provide a valuable statistical sample of around 350 systems. Currently only 96 known lenses satisfy these criteria, yet promisingly, our unWISE modelling technique is able to recover all of these with simple WISE-\Gaia colour cuts that remove $\sim$80 per cent of previously followed-up contaminants. Finally, we provide an online database of known lenses, quasar pairs, and contaminant systems.     

\end{abstract}




\section{Introduction} 

The Pan-STARRS $3\pi$ survey \citep[PS, ][]{chambers2016} is expected to have already imaged $\sim$2000 lensed quasars with images brighter than $i=22$ and separated by more than 0.67 arcseconds \citep{oguri2010}. Considering just the extragalactic (|b|>20) sky of PS (68 per cent of the sky), the number drops to $\sim$1350. However, we currently know of only 163\footnote{https://www.ast.cam.ac.uk/ioa/research/lensedquasars/} lenses satisfying the same criteria. Many of the missing lenses will have their multiple quasar images approximately separated by the typical FWHM of the Pan-STARRS PSF. This blending, coupled with the variable source colour, lensing galaxy brightness, and image configuration, creates a plethora of possible pixel patterns, many of which are easily confounded with vastly outnumbering systems such as quasars projected near stars, star forming galaxies, or even physically distinct quasar pairs. A potential solution to finding all lensed quasars would be to use similar variability between images, as proposed by \citet{kochanek2006}. Indeed, the \citet{oguri2010} estimate expects detection of the brightest lensed quasars by using the PS yearly stacks rather than the full survey depth. Time-resolved data are currently unavailable however, and without the auxiliary information from variability, contaminant systems cannot be readily removed.

We therefore aim to use \Gaia data to help remove contaminants, at the expense of a brighter image threshold. \textit{Gaia}'s excellent resolution \citep[FWHM$\approx$0.1arcsec,][]{fabricius2016} and targeting of point sources \citep{prusti2016,brown2016}, makes it an excellent discovery survey for bright lensed quasars \citep{surdej2002, jackson2009,finet2016}. During its 5 year mission, \Gaia will measure auxiliary information including colours and proper motions for all detections. Proper motions are evidently an effective way to remove contaminant systems containing stars \citep[e.g.][]{heintz2018}. 

The first \Gaia data release, on 14 September 2016, catalogued known lensed quasars within its brightness threshold, but it rarely catalogued all images \citep{lemon2017, ducourant2018a}. The second \Gaia data release \citep[GDR2,][]{gaia2018}, on 25 April 2018, has a marked improvement in detecting closely separated images, down to 0.4 arcseconds \citep{arenou2018}. Accordingly, there is a more complete detection of all images of known lensed quasars \citep{ducourant2018b}.

Given \textit{Gaia}'s whole-sky, high-resolution catalogue, several methods have already been explored to mine gravitational lenses as efficiently as possible. Several techniques have been suggested and applied using such a catalogue of positions, including: matching multiple \Gaia detections to photometric quasars \citep{agnello2018a,krone-martins2018,lemon2018}, considering \Gaia offsets to SDSS quasar positions \citep{lemon2017}, or by using the spatial positioning of clusters of point sources that are consistent with simple gravitational lens models \citep{delchambre2018, wynne2018}. 

Papers I and II in this series developed and applied techniques to discover lensed quasars based on detections in \textit{Gaia}'s first data release. In this paper we search the \Gaia DR2 database, together with data from WISE, SDSS, and Pan-STARRS, to discover 22 new lensed quasars. The paper is outlined as follows: Section \ref{predictions} considers the number of lenses we expect to find given certain constraints on brightness, separation, and survey area using \Gaia and Pan-STARRS; Section \ref{gaiadata} presents the \Gaia parameters of the known lenses and contaminant systems; Section \ref{search} outlines our main search techniques, with the results from a spectroscopic follow-up presented in Section \ref{results}. Section \ref{discussion} investigates a population analysis of these new lenses and of known lenses in comparison to a complete mock sample. We conclude in Section \ref{conclusion}.

\section{Current Lens Statistics and Lens Database} \label{predictions}
We have motivated a search based on the advent of a new dataset, but it is prudent to consider what searches have already been made in the Pan-STARRS footprint, what biases exist in these searches, and how many lenses are already known. Since we will present a search based on optical detections, we will consider the lens searches carried out in the optical, though several radio lens searches \citep[e.g.][]{king1999, browne2003} have found lenses with bright optical counterparts. To aid our understanding of the current lens statistics, we have compiled a list of all known published lensed quasars, with information on image separation, source redshift, and relevant publications in an online database: https://www.ast.cam.ac.uk/ioa/research/lensedquasars/. This database will be continually updated with newly published lensed quasars.

The most extensive optical survey for lensed quasars was the SDSS quasar lens search \citep[SQLS,][]{oguri2006}, which yielded 62 lensed quasars \citep{inada2012}. This search started from spectroscopically confirmed quasars with 0.6 < $z$ < 2.2 and an SDSS component brighter than $i=19.1$. The redshift cut ensured completeness in this redshift range, while at higher redshifts, objects classified as extended were not completely targetted for SDSS spectroscopic follow-up \citep{richards2002}. We expect most small-separation lensed quasar systems with multiple \Gaia detections (i.e. $G$<20.7 for each image) to be brighter than the SQLS selection limit, particularly when considering the lensing galaxy flux. However, wider-separation lenses with the individual components separately catalogued by SDSS might still remain undiscovered. A similar survey conducted for the SDSS-III BOSS quasar sample discovered 13 lenses \citep{more2016}. Concurrently with the SQLS search, lenses were found using an infrared-excess from 2MASS \citep{ofek2007} or using the better quality, redder imaging  of UKIDSS \citep{jackson2008,jackson2009,jackson2012}. More recent lens discoveries in Pan-STARRS, SDSS, and VST-ATLAS have yielded several new lenses, through photometric selection \citep{schechter2017,schechter2018, agnellovst2017,ostrovski2018, williams2018}, further mining of the SDSS spectroscopic quasar sample \citep{sergeyev2016, shalyapin2018}, catalogued deblending \citep{rusu2018}, or serendipity  \citep{berghea2017,lucey2018}. We therefore expect that lenses with high-redshift sources or bright lensing galaxies, and fainter systems with large separations, to still be plentiful in the SDSS and Pan-STARRS footprints.

To understand the number of lenses that could be mined in the Pan-STARRS footprint with multiple \Gaia detections, we use the mock simulations from \citet{oguri2010}, hereafter OM10, and find which lenses on the sky would readily be detected by \textit{Gaia}. To convert the OM10 $i$-band lensed image magnitudes to \Gaia $G$-band magnitudes, we use the SDSS spectroscopic quasar catalogue cross-matched to \textit{Gaia}, so a $G-i$ relation is determined at each redshift. For each mock lens in the catalogue, a \Gaia $G$-band magnitude is synthesised from the matched catalogue based on the source redshift. In Table \ref{tab:om10pred}, we provide the expected number of lensed quasars with \Gaia detections (2 for doubles and at least 3 for quads) in the 21039 square degrees of |b|>20 sky of Pan-STARRS for several image separations. As there is scatter in the $G-i$ relation, the $G-i$ conversion is repeated for each mock lens 100 times and thresholds re-applied. This provides the uncertainties in Table \ref{tab:om10pred}. We also provide the numbers of \textit{known} lenses in the |b|>20 Pan-STARRS footprint that have measured \Gaia detections of G<20.7.

\begin{table}
	\centering
	\caption{Predicted numbers of lensed quasars. The uncertainties come from the $G-i$ band conversion and shot noise from the size of the catalogue. Following the convention of OM10, we include doubles that have both images brighter than the \Gaia threshold, and quads with at least 3 images brighter than this threshold.}
	\label{tab:om10pred}
	\begin{tabular}{ccccc}
		\hline
		 & Im. Sep. (\arcsec) & All & Quads & Doubles \\
         \hline
        PS 3$\pi$ & >0.5 & 352$\pm$9 & 66$\pm$4 & 286$\pm$8 \\
         $G_{2,3}$ < 20.7 & >1.0 & 242$\pm$8 & 46$\pm$3 & 195$\pm$7 \\
         |b|>20 & >1.5 & 145$\pm$6 & 26$\pm$2 & 119$\pm$5 \\
         & >2.0 & 81$\pm$4 & 13$\pm$2 & 68$\pm$4 \\
         & >2.5 & 41$\pm$3 & 6$\pm$1 & 35$\pm$3 \\
		\hline
       	known lenses & >0.5 & 94 & 13 & 81 \\
         in PS & >1.0 & 89 & 13 & 76 \\
        $G_{2,3}$ < 20.7 & >1.5 & 61 & 11 & 50 \\
        |b|>20 & >2.0 & 31 & 8 & 23 \\
        & >2.5 & 17 & 6 & 11\\
        \hline
       
	\end{tabular}
\end{table}

Considering the case of image separations above 1 arcsecond, there are 46/195 expected quads/doubles, though we only know of 13/76 such systems. \Gaia is very complete for detecting both images of doubles when it should (i.e., when both images are brighter than $G\sim20.7$; see Section \ref{detectionrate}), though this might not be true for quads given the increase in local crowding. In both the mocks and known lens numbers, we have included all quads satisfying the maximum image separation, however we have not removed those with minimum image separations less than 0.4 arcseconds. This could partly explain the large discrepancy at the smallest image separations.

\section{\Gaia DR2 data} \label{gaiadata}
Before searching the \Gaia DR2 catalogue for lensed quasars, we investigate how the \Gaia parameters can be used to remove common contaminants. As part of this investigation and to aid future lens searches, we have compiled a list of all published, spectroscopically followed-up lensed quasar candidates. This is readily accessible on our online database, and is intended to both prevent repeated observations, and to allow tests of future lens-finding techniques' selection purity.

\subsection{Detection Rate} \label{detectionrate}
Table \ref{tab:gaiacount} shows the number of images detected by \Gaia for the known lensed quasars in \Gaia DR1 and DR2. We split the sample into doubles and quads, with the latter including all systems with more than 2 images. Only one quad lensed quasar had four images detected in DR1, SDSSJ1004+4112, a quasar lensed by a cluster with images separated by $\sim$15 arcseconds \citep{inada2003}. There is a dramatic improvement for lensed quasar image detection in DR2, with 3 times more doubles having both images detected. The 24 doubles that have only one image detected are due to the fainter image not meeting the detection threshold. The triple system, APM08279+5255, is the only system of the doubles and triples with two images bright enough for \textit{Gaia} to detect, yet with only one detection \citep{irwin1998}. This is likely due to the small separation of the system---0.38 arcseconds. The improved completeness generally applies to the quads as well. The small number of quads with only one image detected in DR2 demonstrates not only how most quads have at least two bright images of similar flux (either in folds, crosses or cusp configurations), but also \textit{Gaia}'s increased detection rate in crowded regions with DR2.

\begin{table}
	\centering
	\caption{Number of lensed quasars with N \Gaia detections in \Gaia DR1 and DR2, from a total sample of 206 lenses---52 quads, 2 triples and 152 doubles. We include triply imaged systems in the quads column.}
	\label{tab:gaiacount}
	\begin{tabular}{ccccccc}
		\hline
        & \multicolumn{3}{c}{\Gaia DR1} & \multicolumn{3}{c}{\Gaia DR2} \\
		& All & Quads & Doubles & All & Quads & Doubles \\
         \hline
       	$N$=0 & 45 & 19 & 26 & 32 & 14 & 18 \\
        $N$=1 & 105 & 14 & 91 & 27 & 3 & 24 \\
        $N$=2 & 50 & 13 & 37 & 122 & 10 & 112 \\
        $N$=3 & 5 & 5 & --- & 12 & 12 & ---\\
        $N\geq$4 & 1 & 1 & --- & 13 & 13 & --- \\
		\hline
	\end{tabular}
\end{table}

\subsection{Proper Motions}
\textit{Gaia}'s precise, multi-epoch photometry also enables proper motions to be measured, and these can be very effective at removing stellar contaminants. Cross-matching SDSS spectroscopic quasars and stars to \textit{Gaia}, all bright ($G$<20) isolated quasars and stars have catalogued proper motions, while only 84 per cent of lensed quasar images, also with $G$<20, have proper motions. This must be due to the crowding of nearby detections and overlapping windows not yet being processed \citep{riello2018}. For all fainter objects ($G$>20) this percentage quickly drops to around 50 per cent.

\begin{figure}
	\includegraphics[width=0.5\textwidth]{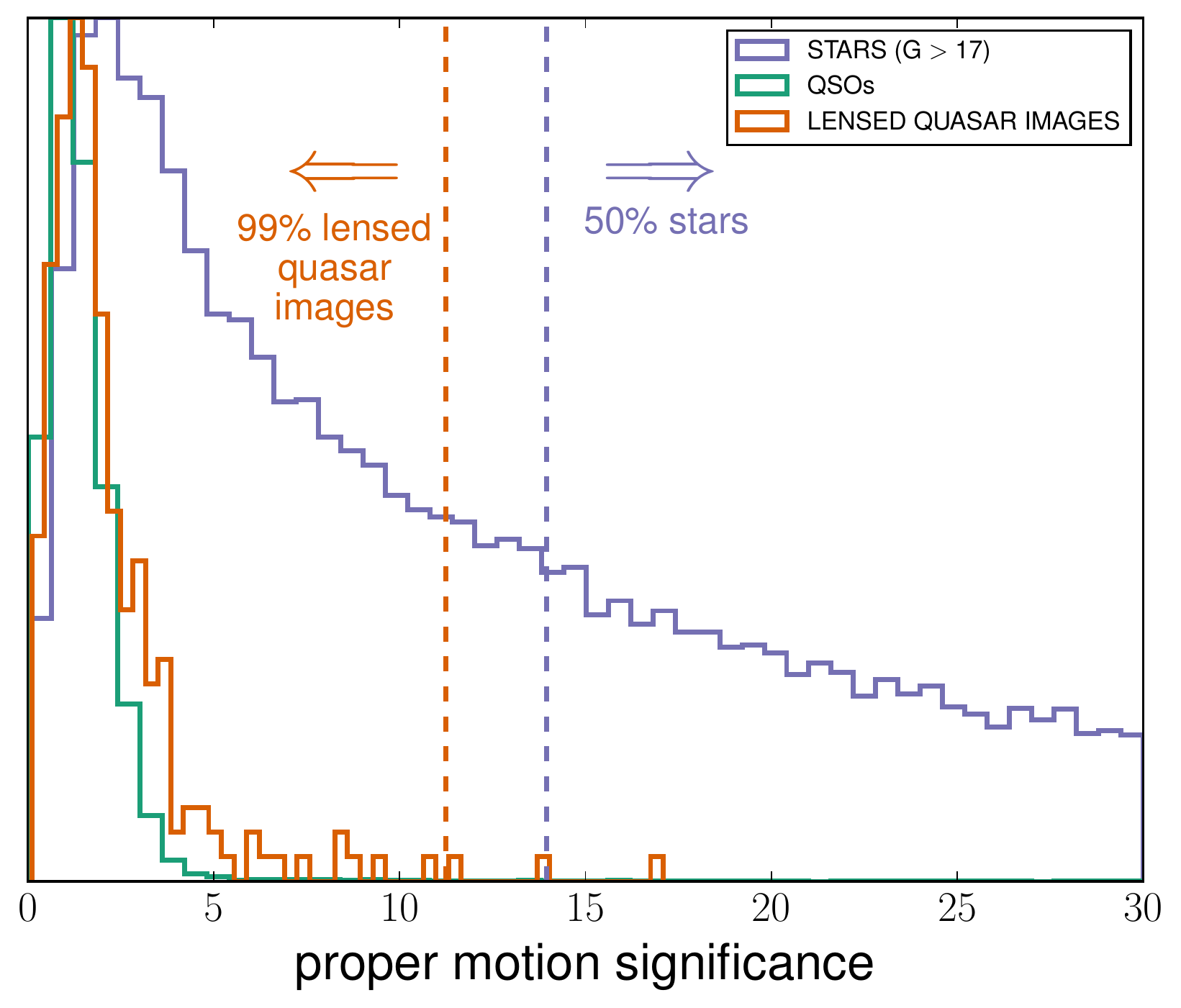}
    \caption{A histogram of proper motion significances for spectroscopic stars fainter than $G=17$, spectroscopic quasars, and lensed quasar images. The histograms are scaled to the same peak values.}
    \label{fig:pms}
\end{figure}

Proper motions alone do not help without taking their uncertainties into account. We therefore define the proper motion significance (PMSIG) as:
\newcommand*\rfrac[2]{{}^{#1}\!/_{#2}}

\begin{equation}
\textrm{PMSIG}={\left[{\left(\frac{\textrm{pmra}}{\textrm{pmra}\_\textrm{error}}\right)}^2 + {\left(\frac{\textrm{pmdec}}{\textrm{pmdec}\_\textrm{error}}\right)}^2\right]}^{\rfrac{1}{2}}
\end{equation}
where pmra, pmra\_error, pmdec, and pmdec\_error are obtained from the \Gaia DR2 catalogue. The distribution of PMSIG for lensed quasar images, isolated quasars, and stars is shown in Figure \ref{fig:pms}. There are 4 lensed quasar images with a PMSIG value above 10: QJ0158-4325 \citep{morgan1999}, DESJ0405-3308 \citep{anguita2018}, RXJ0911+0551 \citep{bade1997}, and SDSSJ1330+1810 \citep{oguri2008}. The latter three cases are compact quads with nearby collections of quasar images, perhaps leading to \Gaia mis-assigning the images at each epoch due to the required binning of a 2.1 arcsecond-wide window perpendicular to the scan direction. However, for the doubly imaged lens, QJ0158-4325, the brighter image ($G$=17.60) has a PMSIG of 13.8 and a small astrometric excess noise of 0.7mas (see Section \ref{aen}), with the other image also being detected by \textit{Gaia} 1.22 arcsec away. It is unclear what has caused this image's large PMSIG.

\subsection{Astrometric Excess Noise} \label{aen}
A major contaminant for lens searches relying on photometric quasar selection is compact star-forming galaxies \citep[e.g.][]{treu2018}. Upon inspection of ground-based data, a pair of star-forming galaxies can appear consistent with PSFs and worthy of spectroscopic follow-up. \Gaia provides a way to remove such systems, through the astrometric excess noise (hereafter AEN) parameter \citep[e.g.][]{belokurov2016,koposov2017}. \Gaia DR1 showed that the AEN could separate intermediate-to-high redshift quasars from star-forming galaxies via a simple cut. In the vast majority of cases this also kept all lensed quasar images \citep{lemon2017}.

\begin{figure}
	\includegraphics[width=0.5\textwidth]{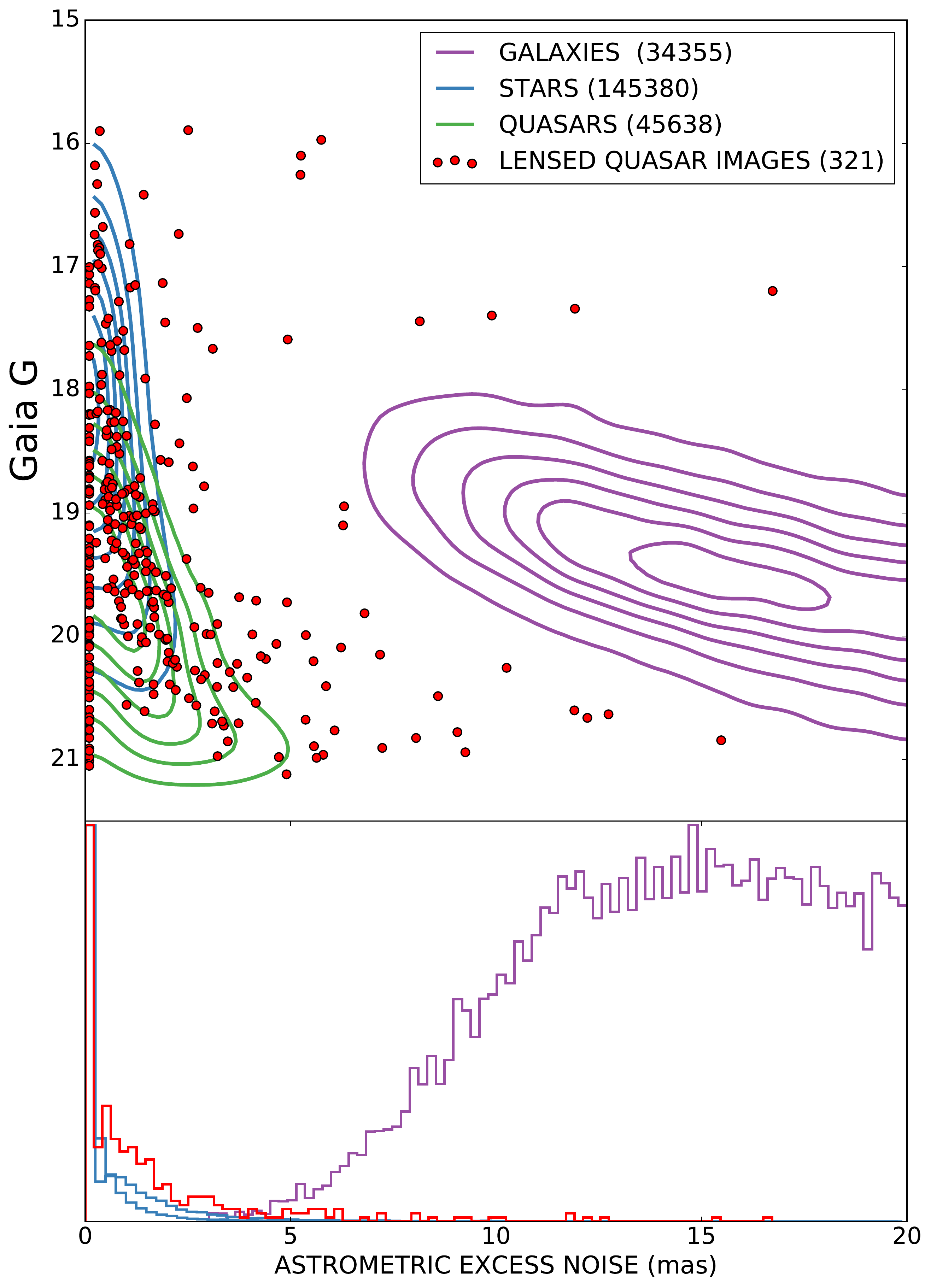}
    \caption{\Gaia $G$ against AEN for SDSS DR12 spectroscopically confirmed stars, galaxies, and quasars. 321 lensed quasar images detected by \Gaia are also overlaid as red dots. They are clearly separated from the galaxy locus.}
    \label{fig:AEN}
\end{figure}

We repeat the \Gaia DR1 AEN comparison from \citet{lemon2017} for \Gaia DR2. From our list of known lensed quasars, there are 321 lensed quasar images with \Gaia detections. Figure \ref{fig:AEN} shows \Gaia magnitude against AEN for each of these images, and also SDSS spectroscopic quasars, galaxies, and stars. All lensed quasar images clearly avoid the galaxy locus in the \Gaia $G$-AEN plane. When only considering AEN, several lens images have galaxy-like AEN values.

The 7 quasar images with AEN>10 fall into three categories. PSJ0840+3550 \citep{lemon2018} and WGD2038-4008 \citep{agnello2018a} have their large AEN image very close to a bright extended galaxy; WFI2026 \citep[a 0.33 arcsecond separation pair,][]{morgan2004} and SDSSJ1640+1932 \citep[a 0.72 arcsecond separation pair, with both images catalogued by \textit{Gaia},][]{wang2017} have very small separation image pairs; and H1413+117 \citep{surdej1988} and DESJ0405-3308 \citep{anguita2018} are compact quads. The first two examples are likely to remain with a large AEN after subsequent \Gaia data releases, though the remainder may improve as further scans at different angles are able to pin down each image. Surprisingly, the two detections in the Einstein cross, Q2237+030, have AENs of 2.2 and 3.0 in spite of all four images being embedded in the centre of the bulge of a relatively local ($z = 0.04$) galaxy.

A search using an AEN cut based on separation would help recover the compact quad systems and still remove wider-separation galaxy pairs. We however only use the AEN to prioritise candidates for most of our searches. Indeed, \Gaia detections could be due to the lensing galaxy, for which a high AEN is expected (see Section \ref{J1524} for such an example). 

\subsection{Removing crowded regions} \label{gaiadensity}
All-sky lens searches must perform a cut to remove very high density regions on the sky. This is usually done by some cut on galactic latitude \citep[e.g.][]{krone-martins2018} around |b|>15 or higher. While this effectively removes the galactic centre, it does not remove star clusters, extended star-forming regions, or the Magellanic clouds, and treats differently crowded areas of sky equally. \Gaia by default allows us to remove regions where the concentration of contaminants (stars near quasars) becomes unmanageable. We do this following \citet{lemon2018} by defining the local \Gaia detection density per square degree within 100 arcseconds of a particular target. A search-dependent cut is then placed on this density. For the Pan-STARRS footprint ($\sim$30832 square degrees), after a density cut of 20000 detections per square degree, the area drops to 22094 (i.e. 72 per cent). For density cuts of 30000, 50000, and 100000 detections per square degrees, the remaining areas are 24336 (79 per cent), 26939 (87 per cent), and 28995 square degrees (94 per cent) respectively.

\section{Lens Selection} \label{search}
In this section, we present three searches for lensed quasars. For each search we use different cuts on number of \Gaia detections, PMSIG, local \Gaia density, and AEN.

\subsection{Multiple \Gaia detections around quasar catalogues}
\subsubsection{WISE}
\begin{figure}
	\includegraphics[width=0.5\textwidth]{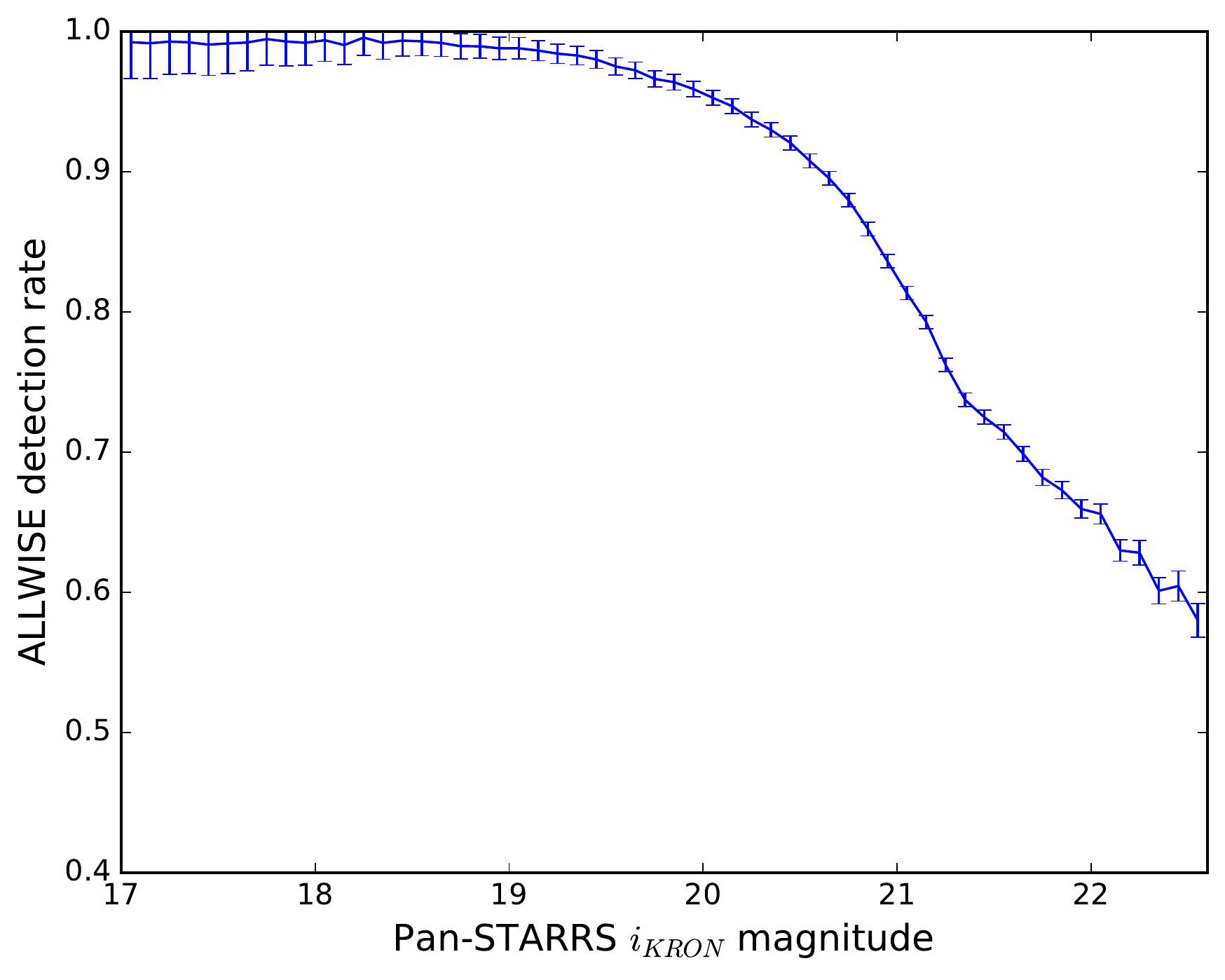}
    \caption{ALLWISE catalogue detection rate of Milliquas quasars (|b|>20), as a function of Pan-STARRS $i_{KRON}$ magnitude. Lensed quasars with two or more Gaia detections ($G$<20.7) will have a total $G$-band (and approximately $i$-band) magnitude less than 19.9.}
    \label{fig:wisedetrate}
\end{figure}

The Wide-Field Infrared Survey Explorer \citep[WISE, ][]{wright2010} provided a full-sky survey at wavelengths of 3.4, 4.6, 12, and 22 microns (W1, W2, W3 and W4). This photometry has since been used as an efficient way to identify pure samples of quasars \citep[e.g.][]{dipompeo2015}. Figure \ref{fig:wisedetrate} shows the ALLWISE detection rate of |b|>20 Milliquas quasars \citep[][see Section \ref{milliquas}]{flesch2015} against Pan-STARRS $i$-band magnitude. The Pan-STARRS-to-ALLWISE crossmatch was made within 4 arcseconds. At optical magnitudes less than 19.5--roughly the faintest magnitude we expect for the sum of quasar images each detected by \textit{Gaia}, and a lensing galaxy detected in Pan-STARRS--nearly all quasars are detected. Such a catalogue is ideal for searching for lensed quasars with multiple Gaia detections; however, we note that a small percentage of bright quasars are not detected due to blending with nearby bright galaxies or PSF spikes. 

We repeat the search from \citet{lemon2018} for multiple \Gaia detections around red ALLWISE detections \citep{mainzer2011}. We visually inspect all sets of 2, 3, and 4 \Gaia detections separated by less than 4.5 arcseconds and all within 4.5 arcseconds of an ALLWISE source satisfying W1$-$W2>0.2, 0.3, and 0.4 respectively, and W1<15.5. Constraints on PMSIG and local stellar density are varied for each selection and summarised in Table \ref{tab:selection}. These cuts are as loose as possible while still keeping the number of systems inspected manageable. 

\subsubsection{Milliquas} \label{milliquas}
The quasar catalogue from \citet{flesch2015} contains nearly 2 million high-confidence quasars\footnote{http://quasars.org/milliquas.htm}. While this catalogue has a large overlap with the WISE colour-selected sample, it includes quasars detected in the X-ray and radio. At these wavelengths a lensed quasar with a bright lensing galaxy will still be selected, whereas in optical- and infrared-selected catalogues, the galaxy can shift the system away from the quasar locus and hence remove it from the selection. For this catalogue we require 2 \Gaia detections both within 4.5 arcseconds of the catalogue position and separated from each other by less than 4.5 arcseconds. This yields 14046 candidates along with a density cut of 20000 and PMSIG less than 12.

\subsection{Modelling unWISE pixels}
\begin{figure*}
\mbox{\subfigure{\includegraphics[width=0.5\textwidth]{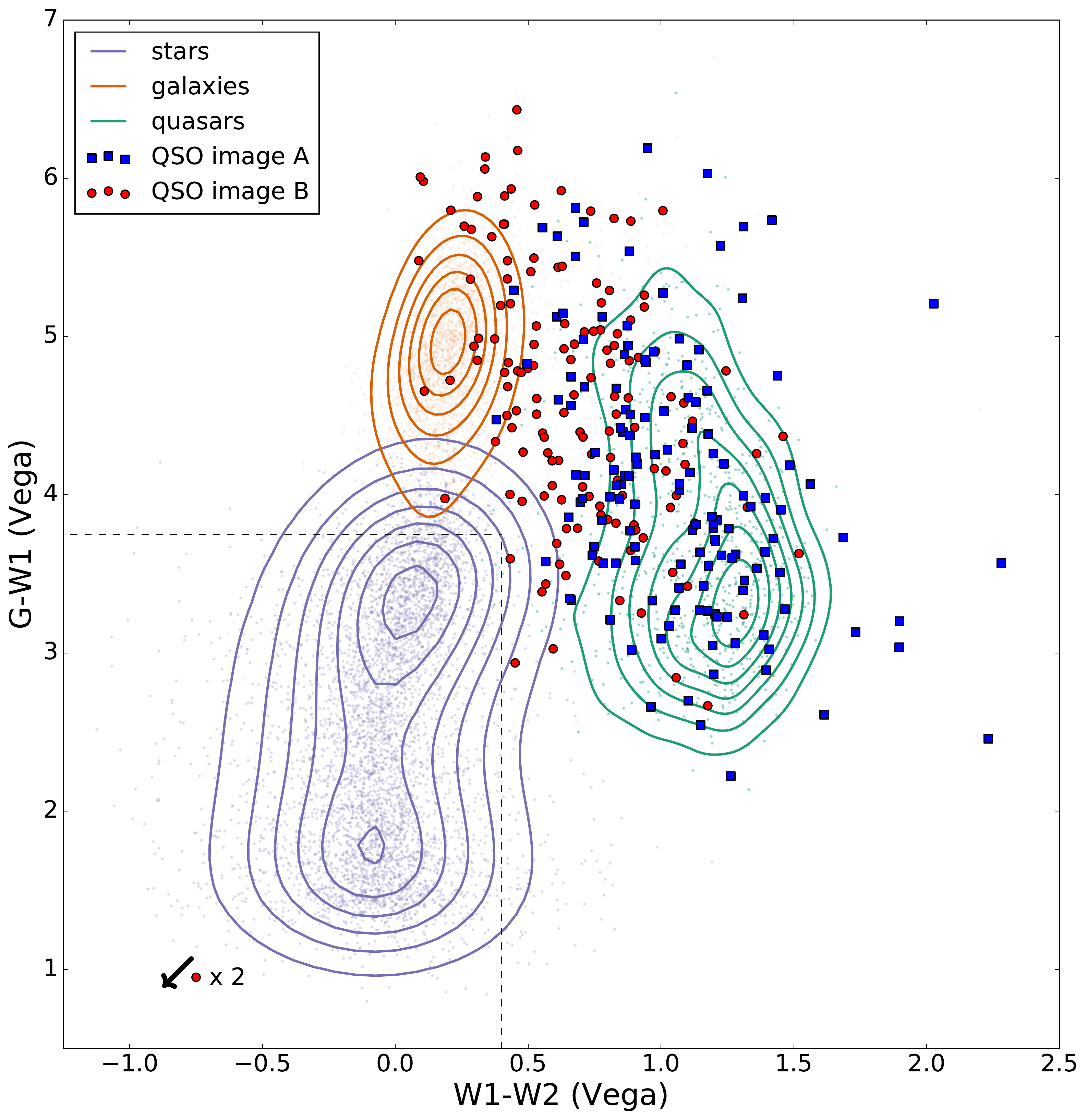}}\quad
\subfigure{\includegraphics[width=0.5\textwidth]{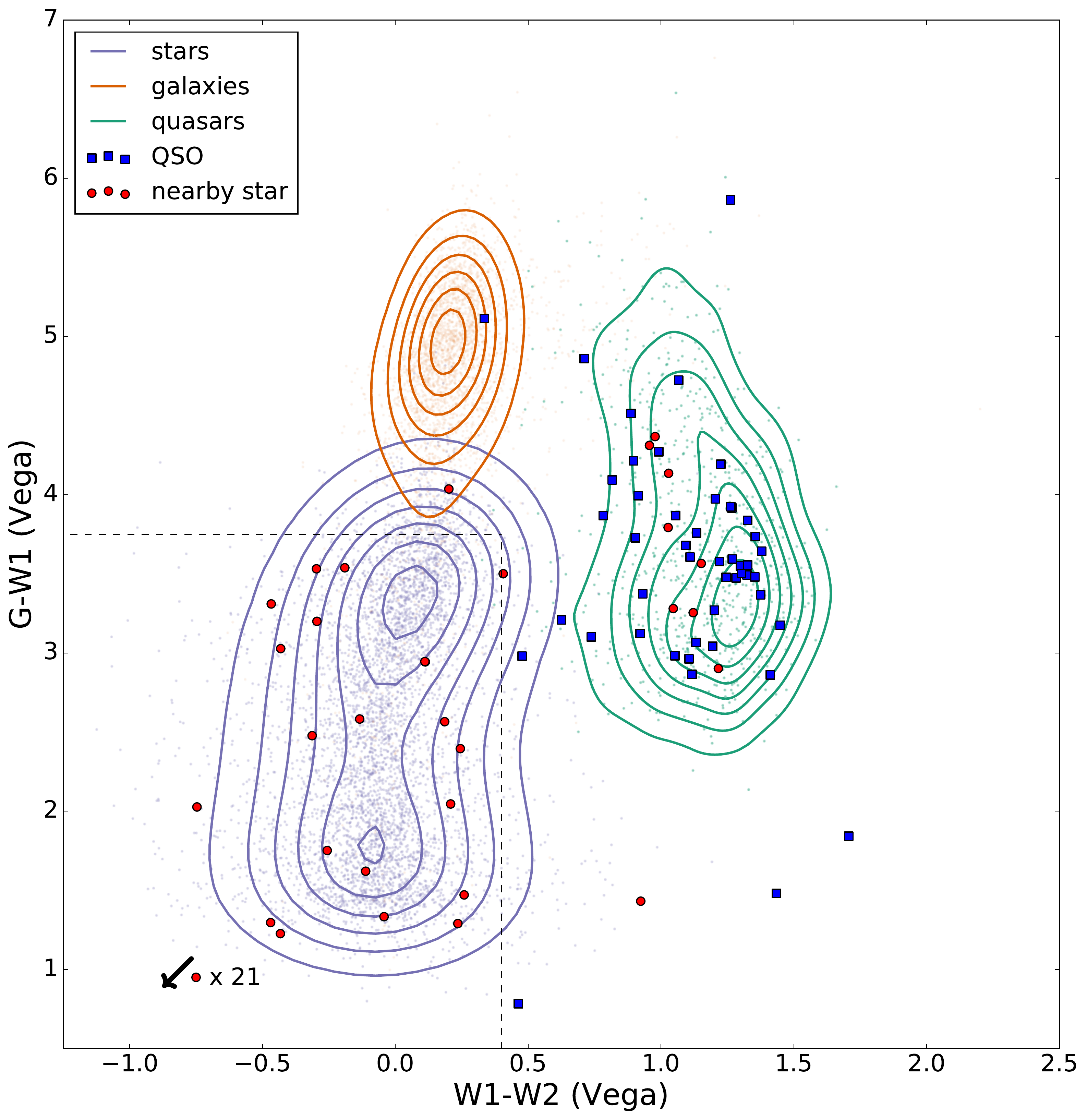}}}
    \caption{\Gaia-WISE colour plots. \textit{Left}: modelled unWISE W1,W2 and \Gaia $G$ colours for all 147 known lensed quasars with two or more \Gaia detections. When more than two images are present, the two with the reddest W1$-$W2 points are plotted. Red circles show the less red component of this pair, i.e. W1-W2(A)$>$W1-W2(B). \textit{Right}: Modelled unWISE W1,W2 and \Gaia $G$ colours for 52 spectroscopically confirmed quasar+star contaminant systems. The stellar components' colours are plotted as red circles. A cut of W1$-$W2<0.4 and $G-$W1<3.75 removed 41/52 contaminant systems, while keeping 145/147 lensed quasars.}
    \label{fig:gaiawise}
\end{figure*}

The FWHM of the WISE PSF in the W1 and W2 bands is $\sim$ 6 arcseconds. While a quasar and star separated by a few arcseconds will be blended and catalogued as one object, the WISE pixels can be modelled if the PSF and positions of the components are known, as has been demonstrated with SDSS data \citep{lang2016} using a set of unblurred WISE coadds \citep[unWISE,][]{lang2014}. We apply this technique, using the \Gaia positions and ALLWISE catalogue positions, to the most up-to-date unWISE coadds and PSFs \citep{meisner2017b,meisner2017a,meisner2018}. The data are very well calibrated astrometrically, allowing us to use the \Gaia positions to model the pixels directly without needing to infer a registration between the two datasets. Any modelled cutouts must be large enough to include PSF spikes from nearby stars, but also small enough for the relative component fit to be computationally fast. Therefore, we divide the sky through an equal area HealPix tiling \citep{gorski2005}, resulting in 50331648  70x70 arcsecond tiles across the whole sky. Each tile overlaps neighbouring tiles by 5 arcseconds in order to ensure that lens candidates at the edges of these tiles are not missed. Applying the following process to each of these cutouts resulted in 25129 candidates:

\begin{enumerate}
\item The \Gaia density (as defined in Section \ref{gaiadensity}) is calculated at the tile centre, and the modelling proceeds if this is below 20000 per square degree.
\item If any close pair of \Gaia detections exists on the cutout separated by 5 arcseconds and any PMSIG values are less than 10, the modelling proceeds.
\item For both the W1 and W2 bands, a model is built by placing unWISE PSFs at the positions of all \Gaia detections, and at ALLWISE catalogue detections which are more than 2 arcseconds from a \Gaia detection. A uniform background is added to this model.
\item The fluxes of each PSF and background level are inferred through a non-negative least squares fit to the data weighted by the uncertainty maps. These are then converted into best fit W1 and W2 Vega magnitudes.
\item If for each component of the close pairs, W1<15.5 and either W1$-$W2>0.4 or G$-$W1>3.75, a Pan-STARRS cutout is visually inspected. See Figure \ref{fig:gaiawise} for these colour cuts.
\end{enumerate}

To test this technique's efficiency at recovering known lenses, we extracted W1 and W2 magnitudes for all 147 known lensed quasars with 2 or more \Gaia detections. The \Gaia $G$, W1, W2 colour plot for the reddest two components of all 147 systems is shown in Figure \ref{fig:gaiawise} (without cuts on separation or PMSIG). From the list of previously followed-up systems (Section \ref{gaiadata}), we compiled a list of 127 spectroscopically confirmed quasar+star systems that were identified as potential lensed quasar systems in previous lens searches \citep{hennawi2006,inada2008,inada2010,inada2012, more2016, lemon2018, williams2018}. 52 of these systems have two \Gaia detections. We plot their modelled WISE+\Gaia colours in Figure \ref{fig:gaiawise}, separating the stellar and quasar components.

Only two lens systems have a component with no modelled W1 or W2 flux, both with \Gaia detections separated by less than 1 arcsecond: SDSSJ1640+1932 \citep{wang2017} and SDSSJ0248+1913 \citep[Ostrovksi et al. in prep.,][]{delchambre2018}. However, 21 such examples exist in the quasar+star sample. Using the previously mentioned colour cuts, we are able to remove 41 of the 52 contaminants, while keeping 145 of the 147 lensed quasars. The contaminant systems are already biased to those systems which most resemble lensed quasars, as the search teams thought they merit spectroscopic follow-up. The 11 contaminant systems which evade our colour cut removal are all 1 arcsecond or less separation.

On application, about 1 in 20 tiles from steps (i) and (ii) are subject to unWISE model fitting, while approximately 1 in 300 of these contain systems passing the remaining colour criteria. The main contaminants are galaxy pairs and pairs of \textit{Gaia} sources on bright WISE star streaks. Since there is no required cut on the W1$-$W2 colour of each component, any source for which \textit{Gaia} is not capturing all the flux---for example, PSFs with nearby galaxies---will lead to an inflated G$-$W1 colour, bringing the system into our inspected sample. Figure \ref{fig:gaiawise} shows that our selection efficiency can be improved by applying a strict cut of W1$-$W2>0, without any change to the completeness.

\subsection{LRGs with \Gaia detections}
Quasar-colour-selected lensed quasar searches will be biased away from discovering systems with bright lensing galaxies \citep[see][]{lucey2018}. We therefore aim to complement our quasar-colour-selected search with a search for \Gaia detections around bright galaxies. Firstly, we use the SDSS DR12 spectroscopic galaxy sample (i.e., any spectra with CLASS='GALAXY'). For each galaxy, we require two detections separated by less than 4.5 arcseconds, and each within 4.5 arcseconds of the SDSS position. A second search is also performed requiring just one \Gaia detection between 1 and 3.5 arcseconds from the galaxy. After a cut on the PMSIG of 8 and AEN of 5, the two searches yielded 3056 and 11201 candidates respectively.

Thirdly, we perform a search for bright morphological Pan-STARRS galaxies satisfying $i_{PSF}-i_{KRON}>$0.7, $i_{KRON}<$19, and with at least one \Gaia detection between 1 and 3 arcseconds away. We apply further cuts of PMSIG < 5 and an AEN<5 on the nearest \Gaia detection, yielding 22156 candidates. These search criteria recover previous lenses with bright lensing galaxies and only one \Gaia detection, i.e. SDSSJ0819+5356 \citep{inada2009}, PSJ0123-0455, and PSJ1602+4526 \citep{lemon2018}.

\begin{table}
	\centering
	\caption{Summary of lens selection techniques. For each method, the number of required \Gaia detections, density cut and PMSIG cut is given. All searches also include a \Gaia magnitude threshold of $G$>15. We note the candidates are not mutually exclusive.}
	\label{tab:selection}
	\begin{tabular}{ccccc}
		\hline
        quasar selection & $N_{Gaia}$ & density & PMSIG & N \\
        \hline
        W1$-$W2$>0.2$ & 4 & $<$100000 & $<$18 & 2895 \\
        W1$-$W2$>0.3$ & 3 & $<$50000 & $<$15 & 10274 \\
        W1$-$W2$>0.4$ & 2 & $<$30000 & $<$10 & 13397 \\
        unWISE model & 2 & $<$20000 & $<$12 & 25129 \\
        Milliquas & 2 & $<$20000 & $<$12 & 14046 \\
        Spec. Gals & 1/2 & $<$20000 & $<$12 & 11201/3056 \\
        Morph. Gals & 1 & $<$20000 & $<$5 & 22156 \\
        \hline
	\end{tabular}
\end{table}

\subsection{Final candidate selection}
The total number of candidates selected by each technique is shown in Table \ref{tab:selection}. For each candidate, a Pan-STARRS \textit{gri} colour image is inspected. The majority of candidates are discarded since they are obvious contaminants---they have components with different colours, or the \textit{Gaia} detections correspond to extended galaxies. The most promising candidates, showing nearby PSFs of similar colours, are graded between 1 and 10. The most promising quad candidates with similar unWISE model colours are given a grade of 10. Doubles showing putative lensing galaxies are graded 7-9 depending on component colours, proper motion values, and astrometric excess noise value, while pairs without lensing galaxies were graded 1-6 depending on the same information.

For any WISE-selected candidates, the W1 and W2 images are checked for contamination from nearby stars/galaxies. Finally, the model $G$-W1 and W1-W2 colours are inspected for all remaining candidates, and their grades adjusted according to their relative position to the quasar locus (see Figure \ref{fig:gaiawise}). The candidates with the highest grades are prioritised for follow-up.

\section{Results} \label{results}
\begin{figure*}
	\includegraphics[width=0.95\textwidth]{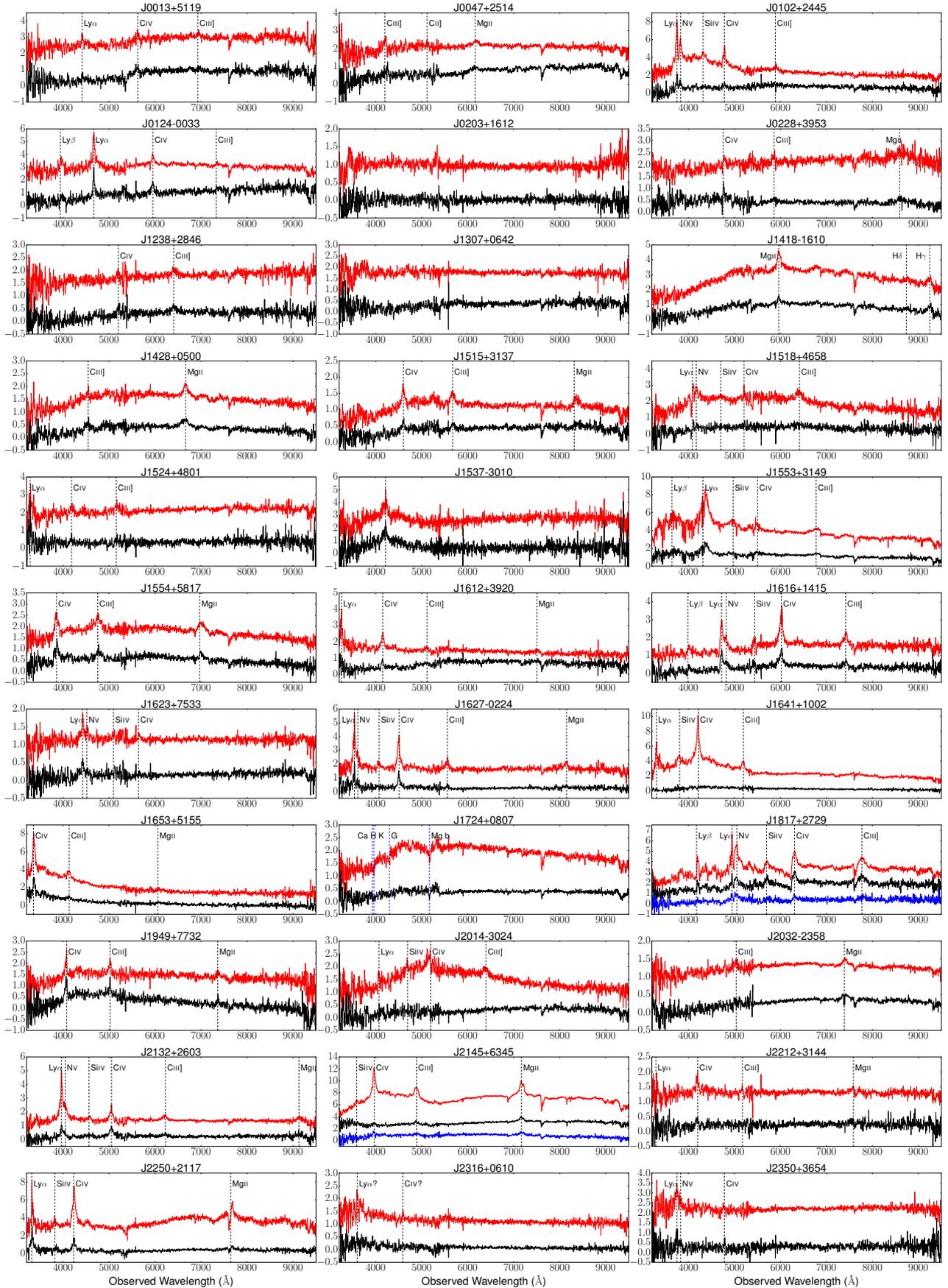}
    \caption{Spectra of all observed candidates. The fluxes of the brighter components (red) have been shifted. J1724+0807 shows absorption lines consistent with $z=0$ stars, while the second components of J1641+1002 and J2014-3024 are featureless.}
    \label{fig:spectra}
\end{figure*}

\begin{figure*}
	\includegraphics[width=\textwidth]{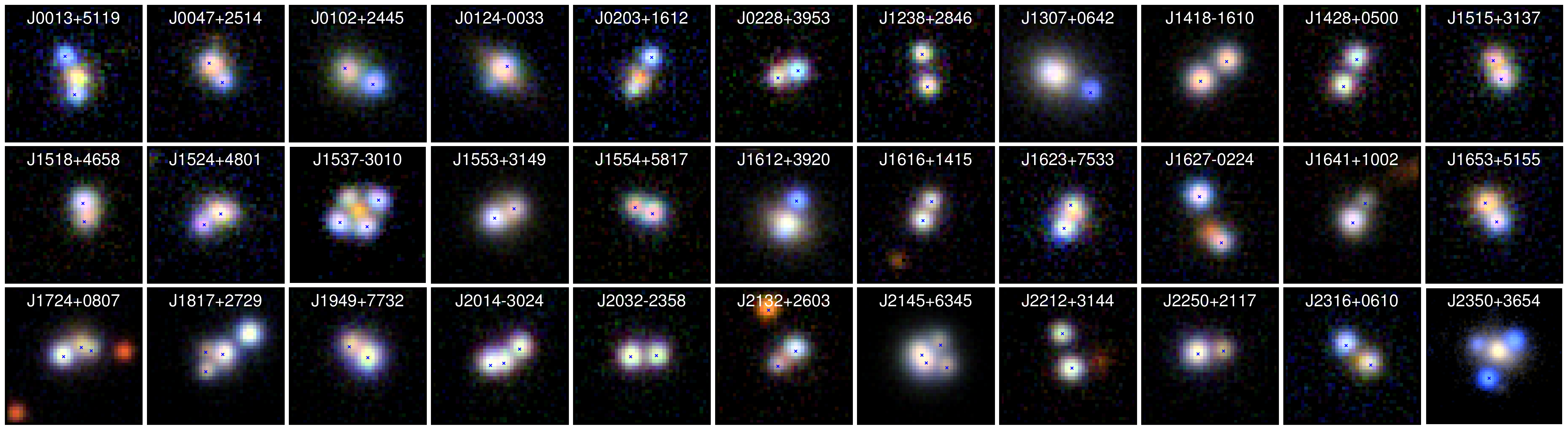}
    \caption{Pan-STARRS \textit{gri} colour images of all observed candidates, except for J1537-3010 which is archival DECam \textit{grz} and J2350+3654 is HSC \textit{gi}. \Gaia detections are overlaid as blue crosses. Cutouts are 10x10 arcseconds in size.}
    \label{fig:colourimages}
\end{figure*}

Spectra of 33 candidates were taken on 11 and 12 July 2018 using ISIS on the William Herschel Telescope in variable seeing and cloud cover conditions. The R158R and R300B gratings were used on the red and blue arms respectively, along with the standard 5300\AA \ dichroic and GG495 second order cut filter in the red arm. This provided a dispersion of 1.81{\AA}pixel$^{-1}$ and 0.86{\AA}pixel$^{-1}$ for the red and blue arms. The component spectra were extracted using a Python-based pipeline which accounts for the CCD response and trace variation based on the standard star Hz 44. The final spectra, shown in Figure \ref{fig:spectra}, were extracted using a Gaussian aperture of varying width centred on each component. The 2D-spectra were also visually inspected to confirm that the spectral features were spatially resolved. Table \ref{tab:observations} provides a summary of all WHT observations, including the \Gaia magnitudes, colour information, and proper motions. Figure \ref{fig:colourimages} shows colour images of all observed systems, with \textit{Gaia} DR2 detections overlaid.

\begin{table*}
	\centering
	\caption{Summary of observations. NIQ=nearly identical quasar, assigned to systems of quasars at the same redshift but without photometric detection of a lensing galaxy. The method of selection is shown for each candidate: WD, WT, WQ = ALLWISE double, triple, quad; uW = unWISE model fitting; MQ = Milliquas; SG1=spectroscopic galaxy with one \textit{Gaia} detection; MG1 = morphological galaxy with one \textit{Gaia} detection.}
	\label{tab:observations}
	\begin{tabular}{lccccccl}
		\hline
		Name & RA & Dec. & Selection & \Gaia G & \Gaia P.M. sig. & Exp. Time & Outcome \\
		\hline
J0013+5119 & 3.34808 & 51.31822 & WD & 20.65, 20.87 & 2.72, 0.73 & 1800s & lens z=2.63 \\
J0047+2514 & 11.94659 & 25.24085 & WD,uW & 20.58, 20.69 & 2.01, - & 1800s & lens z=1.20 \\
J0102+2445 & 15.6965 & 24.7543 & MQ,WD,uW & 19.34, 20.22 & 0.76, - & 1200s & lens z=2.085\\
J0124-0033 & 21.23943 & -0.5533 & MG1 & 20.67 & 1.05 & 2400s & lens z=2.84 \\
J0203+1612 & 30.9977 & 16.20213 & SG1 & 20.76 & - & 1200s & probable lens, \\ &&&&&&& z=0.488, 2.18 \\
J0228+3953 & 37.046244 & 39.88536 & MQ,WD,uW & 20.37, 20.93 & -, - & 1600s & lens z=2.07 \\
J1238+2846 & 189.67778 & 28.78297 & MQ,WD,uW & 20.75, 20.88 & 0.52, - & 2400s & lens z=2.355 \\
J1307+0642 & 196.92947 & 6.70376 & SG1 & 20.16 & 2.44 & 1800s & inconclusive \\
J1418-1610 & 214.57367 & -16.16897 & MQ,WD,uW & 18.46, 19.33 & 1.34, 2.50 & 1200s & NIQ z=1.13 \\
J1428+0500 & 217.23102 & 5.00552 & MQ,WD,uW & 19.89, 19.96 & 1.50, 2.26 & 600s & NIQ z=1.38 \\
J1515+3137 & 228.91601 & 31.62784 & MQ,WD,uW & 19.97, 20.83 & 1.14, - & 700s & lens z=1.97 \\
J1518+4658 & 229.51286 & 46.97113 & WD,uW & 19.82, 21.04 & 2.83, - & 900s & lens? z=2.36 \\
J1524+4801 & 231.12428 & 48.02056 & WD,uW & 20.29, 20.60 & 1.84, - & 1500s & lens z=1.70 \\
J1537-3010 & 234.3556 & -30.17134 & WT & 20.22, 20.32, 20.45 & -, -, - & 2500s & lens z=1.72 \\
J1553+3149 & 238.4092 & 31.82542 & WD,uW & 18.38, 19.43 & 1.42, 2.55 & 600s & lens z=2.55 \\
J1554+5817 & 238.5769 & 58.29635 & MQ,WD,uW & 19.17, 20.01 & 1.13, 1.09 & 450s & NIQ z=1.49 \\
J1612+3920 & 243.05136 & 39.34632 & MG1 & 20.08 & 0.26 & 900s & lens z=1.68 \\
J1616+1415 & 244.1934 & 14.26214 & WD,uW & 19.89, 20.48 & 2.60, 0.51 & 600s & lens z=2.88 \\
J1623+7533 & 245.82049 & 75.55507 & WD,uW & 19.47, 20.01 & 0.30, 2.68 & 1500s & lens z=2.64 \\
J1627-0224 & 246.9594 & -2.40363 & WD,uW & 19.25, 20.08 & 0.99, 0.55 & 600s & lens z=1.91 \\
J1641+1002 & 250.35522 & 10.0484 & MQ,WD,uW & 18.82, 20.95 & 1.48, - & 1200s & z=1.72 quasar + other \\
J1653+5155 & 253.43865 & 51.91796 & WD,uW & 19.94, 20.91 & 1.97, - & 750s & lens z=1.165 \\
J1724+0807 & 261.07672 & 8.12308 & WD,uW & 18.82, 20.26, 20.44 & 9.71, -, - & 600s & stars+galaxy \\
J1817+2729 & 274.37855 & 27.49447 & WQ & 18.93, 20.07, 20.72 & 3.05, 2.56, 0.75 & 1050s & quad z=3.07 \\
J1949+7732 & 297.40117 & 77.54416 & MQ,WD,uW & 18.71, 19.51 & 1.22, 2.68 & 600s & lens z=1.63 \\
J2014-3024 & 303.72582 & -30.41457 & WT & 18.76, 19.09, 19.24 & 4.51, 5.00, - & 450s & z=2.35 quasar+star \\
J2032-2358 & 308.15721 & -23.97286 & MQ,WD,uW & 19.12, 19.26 & 1.23, 0.67 & 800s & NIQ z=1.64 \\
J2132+2603 & 323.0079 & 26.0517 & WT & 19.76, 20.89 & 0.91, - & 600s & lens z=2.26 \\
J2145+6345 & 326.27159 & 63.76145 & WQ & 16.86, 17.26, 18.34, 18.56 & 4.39, 2.86, 1.87, 1.10 & 900s & quad z=1.56 \\
J2212+3144 & 333.03355 & 31.73785 & MQ,WD,uW & 19.28, 19.97 & 1.39, 1.58 & 1200s & lens z=1.71 \\
J2250+2117 & 342.64396 & 21.28988 & MQ,WD,uW & 18.62, 20.27 & 1.93, - & 600s & lens z=1.73 \\
J2316+0610 & 349.13353 & 6.18036 & MQ,WD,uW & 20.05, 20.75 & 1.18, - & 1900s & inconclusive \\ &&&&&&& (lens z=1.96?) \\
J2350+3654 & 357.53143 & 36.90959 & MG1 & 21.12 & - & 3300s & lens z=2.085 \\
		\hline
	\end{tabular}
\end{table*}

\subsection{Modelling} \label{pixelmodelling}
The \textit{grizY} Pan-STARRS images are modelled as in \citet{lemon2018}. For each observed system, a Moffat profile is fit to a nearby star to determine the PSF. All quasar images are fit with PSFs, and galaxies with Sersic profiles convolved with the PSF. The positions and galaxy parameters are inferred using \textit{emcee} \citep{foreman-mackey2013}, while allowing for an astrometric offset between each band. When the nearby star's Moffat profile is not a good fit for the lens system's PSF, the Moffat parameters are inferred simultaneously with the astrometry. The astrometry and photometry for all systems are provided in Appendix A. Figure \ref{fig:psf_subtraction} shows \textit{gri} colour images of each system, with galaxy and PSF positions overlaid, alongside model galaxy-subtracted and model PSF-subtracted \textit{gri} images. 

For each lens we constrain a singular isothermal ellipsoid (SIE) mass model using the inferred galaxy and image positions and their uncertainties. These positions provide 6 constraints for doubles and 10 for quads (excepting J2145+6345, for which no lensing galaxy is detected in the Pan-STARRS data). The model requires a source position, galaxy mass position, and SIE Einstein radius, axis ratio, and position angle, i.e. 7 parameters. We also use the median flux ratios in the \textit{griz} bands, with a 20 per cent uncertainty included to mitigate against microlensing or variations arising from the different light paths. For the quads, flux ratios are also used as constraints, with the uncertainty set at 50 per cent for saddle points, since microlensing is more likely to suppress them \citep{schechter2002}. An external shear is included for the quad mass models. We expect ${\chi}^{2}\sim$ 0 for the double systems, and ${\chi}^{2}\sim$ 4 for the quads (${\chi}^{2}\sim$ 2 for J2145+6345). The best-fit mass model parameters, galaxy light flattenings, and lens model magnifications are given in Table \ref{tab:massmodels}. The ${\chi}^{2}$ contributions from the image positions, lens mass-light misalignment, and flux ratios are also given for each system's best-fit model.

\begin{figure*}
\centering
    \caption{Pan-STARRS modelled pixels. \textit{Left:} \textit{gri} colour image; \textit{middle:} model galaxy subtracted, \textit{right:} model quasar PSFs subtracted. PSF and galaxy positions are overlaid with blue and red crosses respectively. In the case of J1653+5155, the MzLS+BASS DR6 \textit{grz} pixels are used. For J1238+2846 and J1616+1415, the colour images are \textit{grz} and \textit{izY} to show the lensing galaxy more clearly.} 
    \begin{tabular}{lll}
    \includegraphics[scale=0.35]{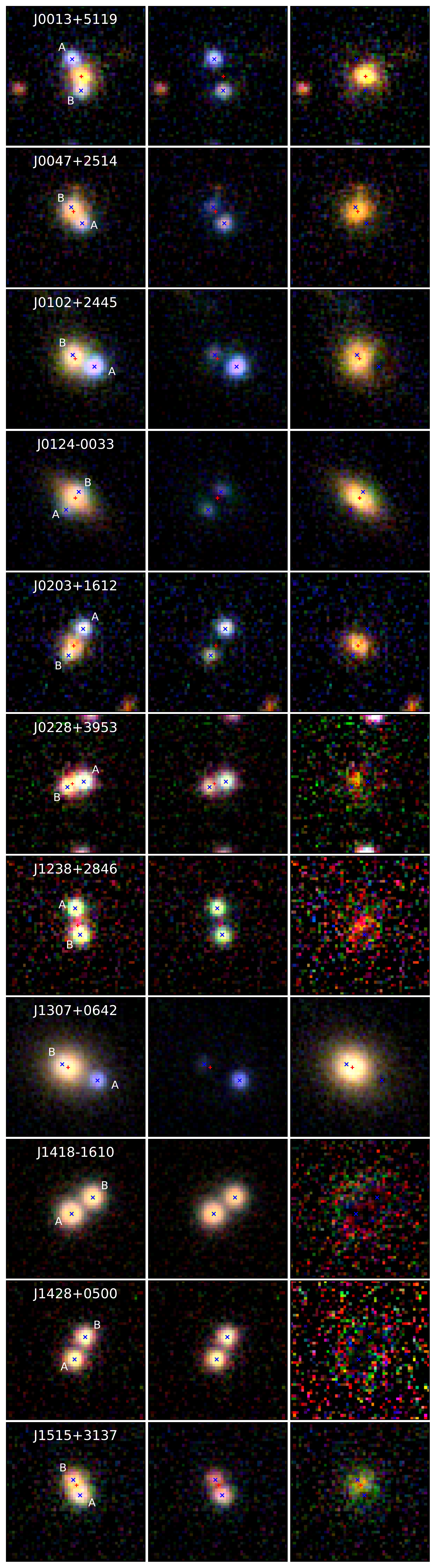}
    \hspace{-2em}
    &
    \includegraphics[scale=0.35]{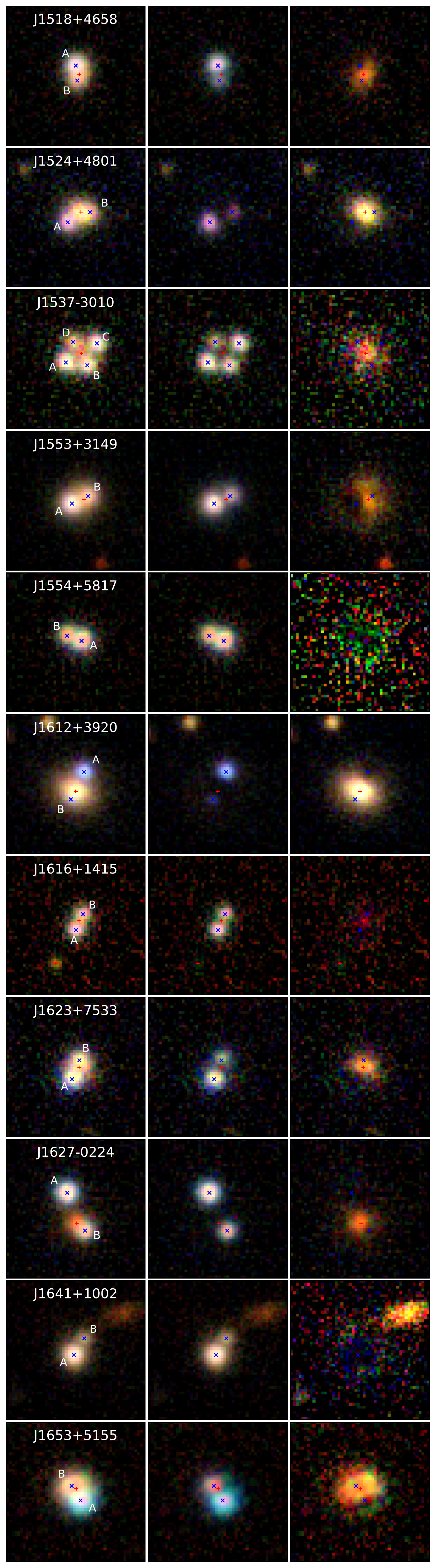}
    \hspace{-2em}
    &
    \includegraphics[scale=0.35]{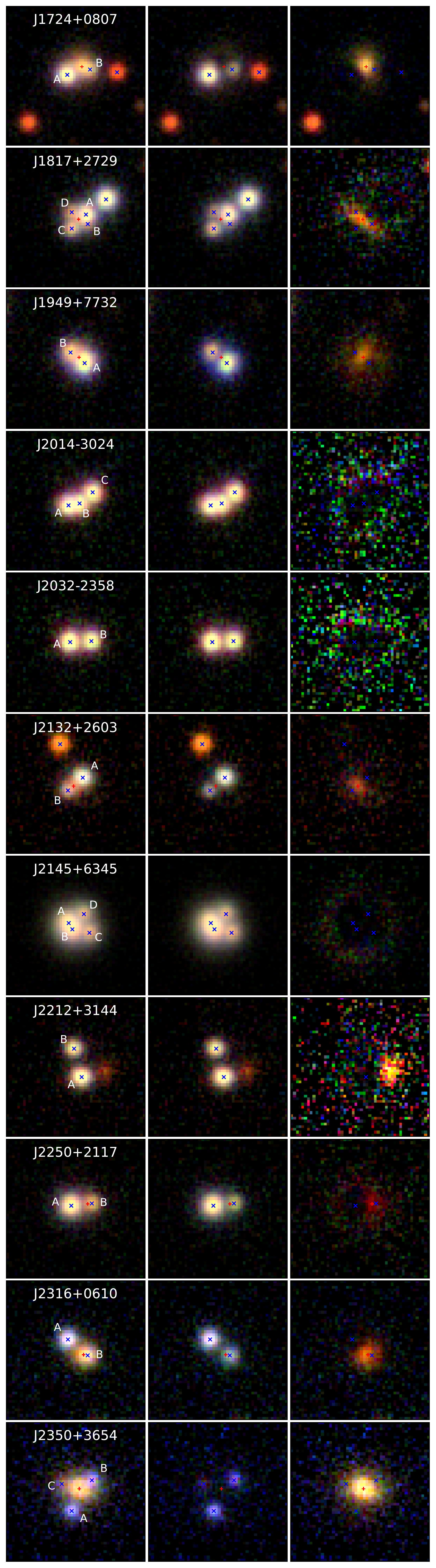}
    \end{tabular}
    \label{fig:psf_subtraction}
\end{figure*}

\begin{table*}
	\centering
	\caption{Median parameter values with 1$\sigma$ uncertainties for mass models of all systems with a lensing configuration. \textit{b}=Einstein radius, \textit{PA}= position angle (East of North), \textit{q} = axis ratio, and $\mu$ = total source magnification.}
	\label{tab:massmodels}
	\begin{tabular}{cccccccc}
		\hline
		name & $b$ ($\arcsec$) & $PA_{SIE}$ &$q_{SIE}$ & $PA_{phot}$ & $q_{phot}$& ${{\chi}^2}_{gal., images, flux}$ & $\mu$  \\
		\hline
        \noalign{\vskip 0.5mm}
        J0013+5119 & $1.51_{0.02}^{0.03}$ & $142.5_{7.4}^{7.4}$ & $0.87_{0.02}^{0.01}$ & $112.1_{4.0}^{4.1}$ & $0.79_{0.03}^{0.03}$ & 0.01, 0.04, 0.00 & $10.7_{0.8}^{0.8}$\\
        \noalign{\vskip 0.5mm}
        J0047+2514 & $0.86_{0.03}^{0.05}$ & $138.3_{13.0}^{19.9}$ & $0.86_{0.13}^{0.09}$ & $171.3_{9.7}^{9.2}$ & $0.89_{0.03}^{0.03}$ & 0.00, 0.07, 0.00 & $3.9_{0.3}^{0.4}$\\
        \noalign{\vskip 0.5mm}
        J0102+2445 & $1.14_{0.03}^{0.06}$ & $1.6_{6.6}^{6.3}$ & $0.50_{0.05}^{0.04}$ & $20.9_{1.1}^{1.1}$ & $0.75_{0.01}^{0.01}$ & 0.00, 0.08, 0.04 & $2.4_{0.1}^{0.1}$\\
        \noalign{\vskip 0.5mm}
        J0124-0033 & $0.95_{0.02}^{0.04}$ & $47.3_{3.3}^{2.5}$ & $0.70_{0.08}^{0.07}$ & $47.9_{0.4}^{0.4}$ & $0.37_{0.01}^{0.01}$ & 0.04, 0.01, 0.00 & $5.8_{0.5}^{0.8}$\\
        \noalign{\vskip 0.5mm}
        J0203+1612 & $1.36_{0.02}^{0.04}$ & $70.4_{36.8}^{83.1}$ & $0.98_{0.03}^{0.02}$ & $56.1_{3.5}^{3.4}$ & $0.66_{0.04}^{0.05}$ & 0.02, 0.06, 0.00 & $7.0_{0.5}^{0.8}$\\
        \noalign{\vskip 0.5mm}
        J0228+3953 & $0.79_{0.01}^{0.03}$ & $162.5_{8.6}^{8.5}$ & $0.80_{0.06}^{0.05}$ & $177.1_{3.7}^{3.2}$ & $0.17_{0.06}^{0.08}$ & 0.03, 0.04, 0.00 & $5.6_{0.9}^{0.9}$\\
        \noalign{\vskip 0.5mm}
        J1238+2846 & $1.19_{0.02}^{0.04}$ & $91.4_{7.8}^{7.0}$ & $0.86_{0.07}^{0.06}$ & $133.5_{10.5}^{11.3}$ & $0.53_{0.15}^{0.16}$ & 0.04, 0.02, 2.21 & $7.0_{1.1}^{2.1}$\\
        \noalign{\vskip 0.5mm}
        J1515+3137 & $0.76_{0.01}^{0.03}$ & $77.0_{20.7}^{18.0}$ & $0.91_{0.08}^{0.05}$ & $146.8_{11.6}^{10.5}$ & $0.81_{0.06}^{0.06}$ & 0.03, 0.04, 0.00 & $6.6_{1.4}^{3.3}$\\
        \noalign{\vskip 0.5mm}
        J1518+4658 & $0.74_{0.01}^{0.03}$ & $152.8_{7.5}^{4.4}$ & $0.76_{0.05}^{0.05}$ & $161.7_{2.7}^{2.7}$ & $0.50_{0.04}^{0.04}$ & 0.06, 0.03, 0.00 & $8.0_{1.1}^{2.8}$\\
        \noalign{\vskip 0.5mm}
        J1524+4801 & $1.19_{0.03}^{0.06}$ & $70.5_{9.1}^{6.5}$ & $0.71_{0.03}^{0.03}$ & $63.5_{1.8}^{2.0}$ & $0.53_{0.03}^{0.02}$ & 0.03, 0.01, 0.02 & $5.4_{0.5}^{0.7}$\\
        \noalign{\vskip 0.5mm}
        J1537-3010 & $1.48_{0.06}^{0.11}$ & $31.8_{1.9}^{1.2}$ & $0.50_{0.09}^{0.35}$ & $33.1_{7.2}^{7.1}$ & $0.74_{0.06}^{0.07}$ & 1.56, 0.41, 2.96 & $7.8_{0.1}^{10.3}$\\
        \noalign{\vskip 0.5mm}
        J1553+3149 & $0.82_{0.01}^{0.02}$ & $157.4_{9.7}^{9.7}$ & $0.80_{0.03}^{0.03}$ & $4.1_{2.2}^{2.2}$ & $0.71_{0.03}^{0.04}$ & 0.05, 0.07, 0.00 & $4.6_{0.4}^{0.3}$\\
        \noalign{\vskip 0.5mm}
        J1612+3920 & $1.37_{0.04}^{0.08}$ & $104.1_{30.4}^{34.5}$ & $0.87_{0.12}^{0.04}$ & $64.2_{0.7}^{0.7}$ & $0.73_{0.01}^{0.01}$ & 0.04, 0.03, 0.00 & $5.2_{0.3}^{0.8}$\\
        \noalign{\vskip 0.5mm}
        J1616+1415 & $0.79_{0.01}^{0.04}$ & $108.8_{28.4}^{27.1}$ & $0.88_{0.10}^{0.08}$ & $160.6_{16.7}^{15.8}$ & $0.77_{0.11}^{0.14}$ & 0.02, 0.05, 2.41 & $8.7_{1.9}^{7.5}$\\
        \noalign{\vskip 0.5mm}
        J1623+7533 & $0.94_{0.02}^{0.04}$ & $29.9_{7.7}^{9.0}$ & $0.76_{0.03}^{0.03}$ & $51.6_{4.4}^{4.4}$ & $0.69_{0.04}^{0.04}$ & 0.03, 0.07, 0.00 & $5.1_{0.4}^{0.4}$\\
        \noalign{\vskip 0.5mm}
        J1627-0224 & $1.91_{0.03}^{0.06}$ & $85.8_{5.1}^{5.3}$ & $0.62_{0.03}^{0.02}$ & $60.6_{20.1}^{13.8}$ & $0.92_{0.04}^{0.04}$ & 0.02, 0.04, 0.00 & $3.3_{0.1}^{0.1}$\\
        \noalign{\vskip 0.5mm}
        J1653+5155 & $0.80_{0.02}^{0.05}$ & $88.4_{7.7}^{8.8}$ & $0.60_{0.05}^{0.04}$ & $93.5_{2.3}^{2.8}$ & $0.81_{0.02}^{0.02}$ & 0.04, 0.01, 0.00 & $3.7_{0.2}^{0.3}$\\
        \noalign{\vskip 0.5mm}
        J1817+2729 & $1.07_{0.03}^{0.04}$ & $54.7_{1.5}^{1.5}$ & $0.22_{0.02}^{0.03}$ & $56.1_{1.4}^{1.9}$ & $0.10_{0.05}^{0.05}$ & 0.21, 0.53, 15.9 & $14.2_{0.9}^{1.9}$\\
        \noalign{\vskip 0.5mm}
        J1949+7732 & $0.80_{0.01}^{0.02}$ & $161.5_{5.6}^{9.2}$ & $0.93_{0.03}^{0.03}$ & $23.8_{66.9}^{44.5}$ & $0.95_{0.05}^{0.03}$ & 0.05, 0.04, 0.00 & $18.2_{3.1}^{6.8}$\\
        \noalign{\vskip 0.5mm}
        J2132+2603 & $0.89_{0.01}^{0.02}$ & $117.4_{27.8}^{18.8}$ & $0.92_{0.04}^{0.05}$ & $20.0_{21.2}^{17.6}$ & $0.82_{0.09}^{0.10}$ & 0.04, 0.06, 0.00 & $7.1_{1.5}^{1.5}$\\
        \noalign{\vskip 0.5mm}
        J2145+6345 & $1.02_{0.00}^{0.04}$ & $166.4_{14.1}^{7.2}$ & $0.86_{0.32}^{0.04}$ & --- & --- & ---, 0.61, 4.67 & $21.0_{0.9}^{0.4}$\\
        \noalign{\vskip 0.5mm}
        J2250+2117 & $0.93_{0.01}^{0.03}$ & $89.9_{56.4}^{61.5}$ & $0.93_{0.07}^{0.05}$ & $72.1_{53.6}^{12.7}$ & $0.88_{0.09}^{0.08}$ & 0.03, 0.06, 0.05 & $3.3_{0.5}^{0.4}$\\
        \noalign{\vskip 0.5mm}
        J2316+0610 & $1.13_{0.02}^{0.03}$ & $126.9_{2.5}^{2.1}$ & $0.29_{0.05}^{0.06}$ & $131.9_{5.1}^{4.8}$ & $0.71_{0.04}^{0.05}$ & 0.03, 0.02, 0.00 & $2.3_{0.1}^{0.2}$\\
        \noalign{\vskip 0.5mm}
        J2350+3654 & $1.73_{0.02}^{0.05}$ & $92.2_{6.2}^{5.7}$ & $0.78_{0.02}^{0.02}$ & $160.3_{13.7}^{15.5}$ & $0.76_{0.10}^{0.13}$ & 0.01, 0.05, 0.04 & $7.1_{0.4}^{0.8}$\\
        \noalign{\vskip 0.5mm}
		\hline
	\end{tabular}
\end{table*}

\begin{table}
	\centering
	\caption{Summary of lensed quasars (image separations are the largest ones for quads).}
	\label{tab:lenssummary}
	\begin{tabular}{cllc}
		\hline
		name & $z_{quasar}$ & sep. & $i_{images}/i_{lens}$ \\
          &  &  (\arcsec) &  \\
		\hline
J0013+5119 & 2.63 & 2.92 & 20.49, 20.54/18.41\\
J0047+2514 & 1.20 & 1.73 & 20.74, 21.74/18.68\\
J0102+2445 & 2.085 & 2.38 & 19.21, 20.69/17.70\\
J0124-0033 & 2.84 & 1.98 & 21.70, 21.11/18.20\\
J0203+1612 & 2.18 & 2.73 & 20.37, 20.97/19.30\\
J0228+3953 & 2.07 & 1.57 & 19.90, 20.15/20.66\\
J1238+2846 & 2.355 & 2.43 & 20.52, 20.47/21.38\\
J1515+3137 & 1.97 & 1.50 & 19.84, 20.42/19.85\\
J1518+4658 & 2.36 & 1.36 & 19.72, 21.08/19.56\\
J1524+4801 & 1.70 & 2.22 & 20.05, 21.00/18.73\\
J1537-3010 & 1.72 & 3.29 & 19.79, 20.00, 19.91, 20.65/19.35\\
J1553+3149 & 2.55 & 1.59 & 18.30, 19.38/18.61\\
J1612+3920 & 1.68 & 2.74 & 20.06, 21.55/17.87\\
J1616+1415 & 2.88 & 1.54 & 19.60, 20.07/20.70\\
J1623+7533 & 2.64 & 1.79 & 19.73, 20.68/19.27\\
J1627-0224 & 1.91 & 3.77 & 19.13, 19.93/19.01\\
J1653+5155 & 1.165 & 1.63 & 20.09, 19.87/19.07\\
J1817+2729 & 3.07 & 1.80 & 18.50, 21.62, 19.61, 20.47/19.92\\
J1949+7732 & 1.63 & 1.59 & 18.82, 19.51/18.98\\
J2132+2603 & 2.26 & 1.77 & 19.44, 20.61/19.75\\
J2145+6345 & 1.56 & 2.07 & 16.71, 16.51, 17.82, 18.08/---\\
J2250+2117 & 1.73 & 1.89 & 18.02, 19.72/19.71\\
J2350+3654 & 2.085 & 3.31 & 20.82, 21.10/18.06\\
		\hline
	\end{tabular}
    
\end{table}

\subsection{Notes on Individual Systems}

\subsubsection{J0203+1612}
\begin{figure}
\includegraphics[width=0.5\textwidth]{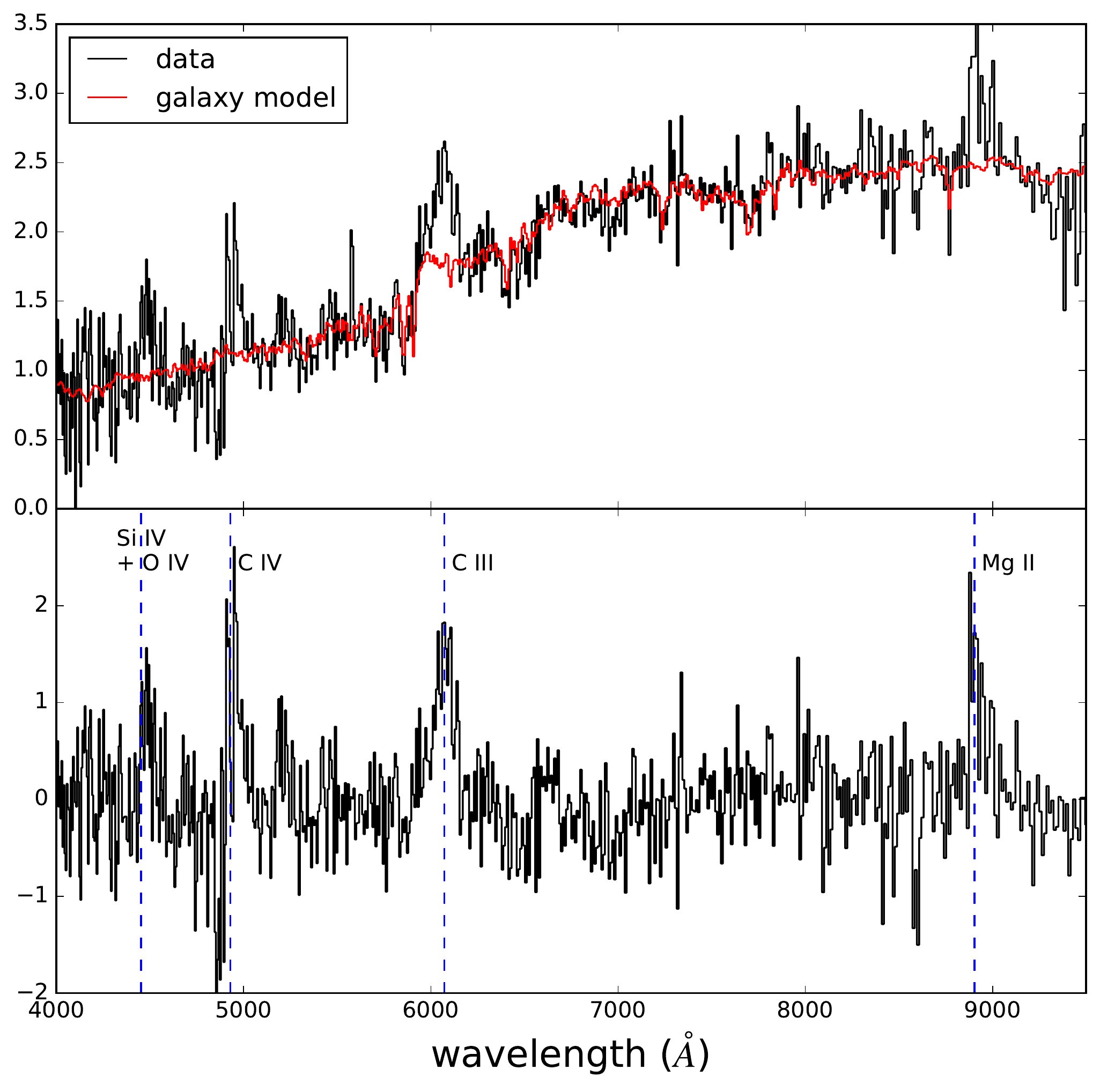}
    \caption{\textit{Top}: BOSS spectrum of J0203+1612, classified as a galaxy with pipeline galaxy model overlaid; \textit{Bottom}: galaxy model subtracted from the data (residual signal-to-noise) clearly showing several quasar emission lines at $z=2.18$}
    \label{fig:J0203+1612}
\end{figure}

This system was selected as a single \Gaia detection near a spectroscopic galaxy. The 1 hour BOSS spectrum reveals a LRG spectrum at $z=0.488$ with obvious quasar emission lines \citep{dawson2013}. Figure \ref{fig:J0203+1612} shows the original BOSS spectrum, best-fit galaxy model spectrum, and the residuals. Four quasar emission lines have been identified at $z=2.18$, with potential broad absorption blueward of the CIV line. We targetted J0203+1612 with the WHT for 1200s however the spectrum (Figure \ref{fig:spectra}) reveals no broad lines, perhaps due to high airmass observations coupled with variable weather conditions. Given the definite presence of a low-redshift galaxy and a high-redshift quasar within the BOSS 2 arcsecond fibre diameter, as well as the imaging showing two point sources either side of the massive galaxy, it is a strong lens candidate. To test this we compare the stellar mass from a composite stellar population synthesis model based on the galaxy's colours and redshift with the lensing mass within the Einstein radius under the assumption of strong gravitational lensing. For the former we used the median colours derived from the pixel modelling (Appendix A), along with the galaxy's spectroscopic redshift. Using the Bayesian stellar population analysis code of \citet{auger2009}, we derive a \textit{total} stellar mass of $\textrm{log}_{10}(M/M_{\odot})=$11.66$\pm$0.05 (11.42$\pm$0.05) for a Salpeter (Chabrier) IMF. In contrast our SIE lens model implies the Einstein mass within the Einstein radius is $\textrm{log}_{10}(M/M_{\odot})\approx$11.64. This lensing mass to stellar mass ratio is typical for samples of strong lenses \citep[e.g.][]{auger2010}, and is on the more conservative side of the gravitational lensing regime since the predicted Einstein radius from the stellar mass is marginally larger than the mass model predicts. On balance, we expect that this is a gravitationally lensed quasar.

\subsubsection{J1238+2846}
The WHT spectra show a shared broad emission line at 6410\AA\ (and perhaps a line at 5210\AA) for each component of this pair. The southern component has a BOSS spectrum clearly revealing much higher signal-to-noise versions of these emission lines, corresponding to CIII and CIV respectively. Interestingly the BOSS spectrum also shows strong Ly$\alpha$ and NV which are not present in our WHT spectrum. This is probably due to the candidate being targetted at the very start of the night at high airmass and with the slit aligned along the two images rather than the parallactic angle, reducing the amount of flux to the blue arm. The lensing galaxy is revealed most readily in the $z$-band and is verified in both the Pan-STARRS and DECaLS $z$-band data.

\subsubsection{J1307+0642}
This system was selected using \textit{Chandra} archival observations and \Gaia (Lemon et al. in prep). The \textit{Chandra} data resolve two X-ray point sources coinciding with two optically blue point sources either side of a bright galaxy (see Figure \ref{fig:psf_subtraction}). The optical spectrum shows hints of an emission line around 4700\AA. A SDSS spectrum reveals the lensing galaxy to be at z=0.23. However, no quasar emission lines are seen upon subtraction of the model spectrum. A deeper spectrum is required to determine the source redshift for this object.

\subsubsection{J1418-1610}
The blue arm reveals no broad emission lines, however as in the case of J1238+2846, it was observed at high airmass. The red arm reveals two prominent emission line at $\sim$5960 and $\sim$9260\AA, suggesting MgII and H$\gamma$ at z=1.13. This is in agreement with the subsequent identification of H$\delta$ and FeII ($\sim$3200\AA) in the stacked spectrum, and also with a UV excess as detected by GALEX. 

\subsubsection{J1524+4801} \label{J1524}
This doubly imaged lens was selected both by the unWISE modelling and the W1$-$W2>0.4 double \Gaia detection search. Figure \ref{fig:colourimages} shows that the two \Gaia detections correspond to a quasar image ($G$=20.29, AEN=1.98), and the lensing galaxy ($G$=20.60, AEN=17.53). Surprisingly the lens galaxy is fainter (\textit{i}=19.56) than most other galaxies with \Gaia detections, suggesting a compact central region. This lens is the first known lensed quasar for which \Gaia has catalogued the lensing galaxy, opening up new search opportunities. Our searches have revealed other candidates with apparent lensing galaxies detected by \textit{Gaia}.

\subsubsection{J1537-3010}
Our WHT spectra reveal that the eastern and southern images share a broad emission line at 4212\AA. A second emission line at 5190\AA \ has been identified with Subaru (Rusu, private communication). This places the source at a redshift of $z=1.72$, with the two lines being CIV and CIII respectively. The SIE+shear mass model provides a good fit, ${\chi}^{2}\sim$5 given 4 degrees of freedom. The majority of this derives from image B being half as bright as the model predicts. The shear has a magnitude of 0.07, 34 degrees East of North. This candidate was also independently selected by \citet{delchambre2018} as GRAL153725327-301017053 and spectroscopically confirmed (Delchambre et al., private communication).

\subsubsection{J1817+2729}
This object was selected by the W1$-$W2>0.2, 4 \Gaia detections search. One of these detections is due to a very nearby star. Given its high stellar density and apparent lack of a lensing galaxy or fourth image, it was given a low follow-up priority. However, we upgraded the priority given its selection by \citet{delchambre2018} as GRAL181730853+272940139, as in the case of J2014-3024. Resolved spectra verify images A, C, and D are at a redshift of $z=3.07$. Our SIE+shear model predicts a highly flattened mass distribution with $q=0.21$ orientated 54 degrees East of North, and a strong shear of 0.27, -34 degrees East of North. The light from the galaxy is also highly elongated, $q=0.1$, and in a remarkably similar direction to the mass (56 degrees East of North). The main contribution to $\chi^2$ is the flux ratio, with image B being 4 times fainter than predicted by our SIE+shear model. This is a saddle point and such images can be strongly demagnified by microlensing \citep{schechter2002}. We note that such a strong shear, orthogonal to the mass distribution likely suggests a limitation in the choice of lensing model. \citet{rusulemon2018} discuss a more intricate mass model, based on deeper Subaru imaging, and suggest that the lensing galaxy is an edge-on disk. The lensing nature has been independently spectroscopically confirmed by Delchambre et al. (private communication).

\subsubsection{J1949+7732}
This lens is an XMM source \citep{pineau2011} and has a source redshift of $z=1.63$. The SDSS data show that the north-eastern component is brightest in all bands, while the Pan-STARRS data clearly show the opposite. The \Gaia magnitudes corroborate the Pan-STARRS flux ratios ($G$=18.71, 19.51 for A and B). Similarly to J0235-2433, it has both images detected by \textit{Gaia}, and so a lightcurve will help reveal the nature of this flux ratio discrepancy, perhaps attributable to an ongoing microlensing event. The modelled unWISE magnitudes are similar for each component: W1,W2=14.94,14.03 for A and W1,W2=15.08,13.97 for B. We note that even though the WISE quasar fluxes should be unaffected by microlensing, they are likely contaminated by flux from the unmodelled lensing galaxy. Our simple mass model predicts a large magnification of $\sim$20, however, this is unreliable since the potentially microlensing-affected flux ratio has been used to constrain the model. 

\subsubsection{J2014-3024}
This system was selected as a \Gaia triple with W1$-$W2>0.3. It was originally given a low priority because of the lack of a fourth image or lensing galaxy and a poor unWISE pixel model--only one component is classified as a quasar. We upgraded the priority given its selection by \citet{delchambre2018}, as in the case of J1817+2729. The spectra of the two eastern PSFs are blended and show features of a quasar at $z=2.35$, while the western PSF has a much redder spectrum, showing no quasar emission lines. Therefore, we can rule out the hypothesis that this is a quadruply imaged lensed quasar, but cannot rule out if the close pair are components of a doubly imaged quasar.

\subsubsection{J2032-2358}
Given the high airmass at which this target was observed, there is little signal from the blue arm. On the red arm, the single shared emission line can be identifed as MgII, since in one component there is the characteristic MgII absorption doublet. A corresponding signal can then be seen in the noisy blue spectra, corresponding to CIV (see Figure \ref{fig:spectra}). This places the sources at $z=1.64$, however no lensing galaxy is detected in the Pan-STARRS images. 

\subsubsection{J2145+6345}
This object was discovered by our W1$-$W2>0.2, 4 \Gaia detections search. Three spectra were extracted for this object, since components A and B could not be resolved. It has a similar configuration and brightness to PG1115+080 \citep{weymann1980}. It lies at |b|=7 in a crowded stellar field with a \Gaia density of 71000 detections per square degree. While such regions are normally excluded from lens searches, proper motions from \Gaia made our search in such regions more manageable by removing many high proper motion contaminant systems. No galaxy is revealed in the PSF subtraction of the Pan-STARRS data. The system is very bright in WISE (W1=12.08 in Vega) and corresponds to a faint ROSAT source \citep{voges1999} with 7 ROSAT photons within 30 arcseconds of the lens. Furthermore, all images are optically bright ($G$=16.86, 17.26, 18.34 and 18.56), making it second only to PSJ0147+4630 \citep{berghea2017} in terms of the brightest fourth image. This makes the system ideal for high-cadence, high signal-to-noise monitoring to determine time delays for cosmography \citep{courbin2018}. Our mass model (SIE+shear) is constrained exclusively by image positions and fluxes, as the lensing galaxy is not detected. It predicts a shear of 0.13, -36 degrees East of North, and an overall magnification of $\sim$21. The fitted galaxy position is at x, y = 0.206, 0.271 arcseconds relative to the astrometry given in Appendix A. 

\subsubsection{J2350+3654}
Figure \ref{fig:J2350} shows a \textit{gi} Hyper-Suprime Cam (HSC) colour image of J2350+3654. A third blue object, C, is seen north-east of the lensing galaxy. Subtraction of the lensing galaxy and quasar images reveals two more sources, one to the West of the lensing galaxy, and one red source west of image A, the brighter image. The $i$-band magnitudes of A, B, and C are 20.82, 21.10, and 21.9. There also exists narrow band (NB515) HSC imaging of this system. The differences in \textit{g-i} and \textit{g-NB515} colours between the three components are: $\Delta (g-i)_{AB, AC}=-0.08, -0.62$ and $\Delta (g-NB515)_{AB, AC}=0.28, -0.09$. The nature of C is still yet to be determined. It might be a third quasar image, but given the single lensing galaxy we would often expect a 4th image. It could also be a secondary source, such as the quasar host galaxy---as in SDSSJ1206+4332 \citep{agnello1206}---or another quasar. C could also be a foreground star. Deeper imaging and spectra of this system are required.

\begin{figure}
\includegraphics[width=0.5\textwidth]{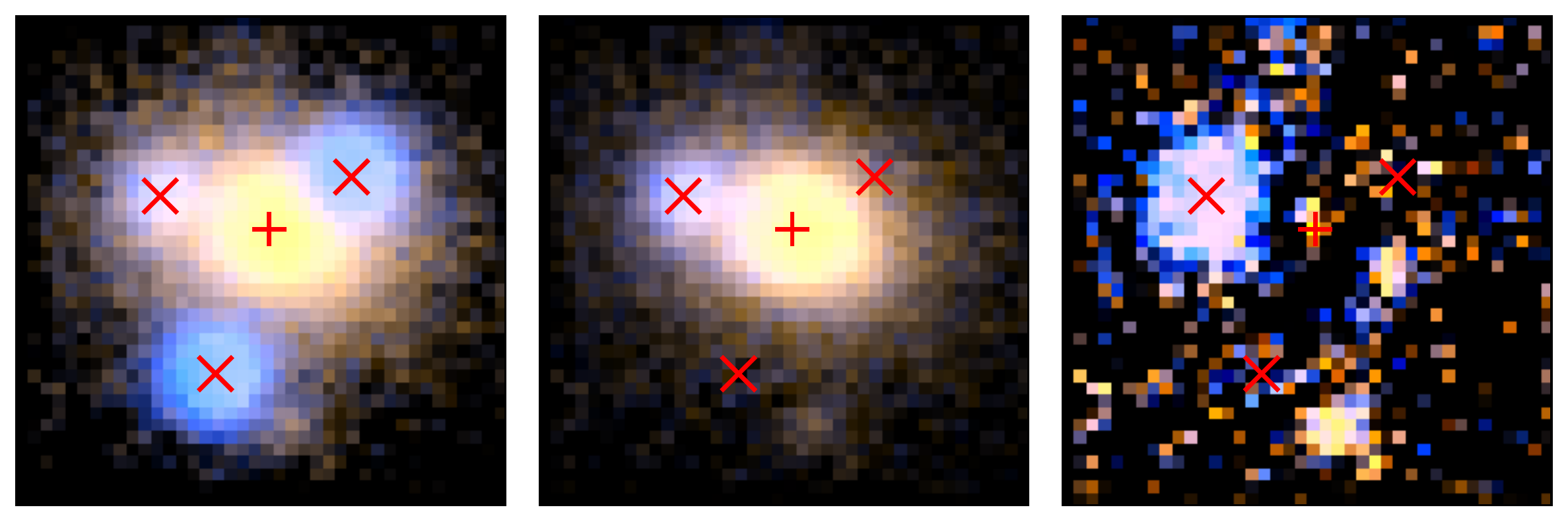}
    \caption{\textit{Left}: A 7x7 arcsecond \textit{gi} HSC colour image of J2350+3654; \textit{middle}: model for the confirmed quasar images (A and B) subtracted from the data; \textit{right}: confirmed quasar and galaxy models subtracted from the data, revealing a bright blue point source and two other sources, one west of the lensing galaxy and one west of image A.}
    \label{fig:J2350}
\end{figure}

\section{Discussion} \label{discussion}
We have observed 33 candidates identified amongst photometric quasar candidates with multiple \textit{Gaia} detections, or a single \textit{Gaia} detection offset from a galaxy. We have found 22 lenses, 2 highly probable lenses and 5 NIQs. 1 candidate is still unclassified, and 3 are contaminant systems. For the two likely lenses, J0203+1612 shows two blue point sources either side of a galaxy, with a spectrum showing blended features of a low-redshift galaxy and a higher-redshift quasar; and J1307+0642 also shows two blue point sources either side of a bright galaxy, each corresponding to a resolved \textit{Chandra} X-ray source. Our low contaminant rate arises from the large number of bright lenses left in the Pan-STARRS footprint (see Table \ref{tab:om10pred}) coupled with a large reduction in typical contaminant systems by using \textit{Gaia}'s catalogue information and unWISE pixel modelling. This modelling has now become particularly useful due to the increase in resolution and detection of small-separation pairs in \Gaia DR2. The limited spectroscopic follow-up time since the second \Gaia data release has allowed us to concentrate on the lens candidates that are most convincing, namely those with bright lensing galaxies. To build a statistical sample from \textit{Gaia}, however, we must ensure that the lensed quasars with fainter lensing galaxies are confirmed. We are confident that these have been selected in our candidate list, since our lens selection techniques recover all previously known lensed quasars with 2 or more \Gaia detections, which includes statistical samples unbiased by bright lensing galaxies (e.g. SQLS).

\begin{figure}
\includegraphics[width=0.5\textwidth]{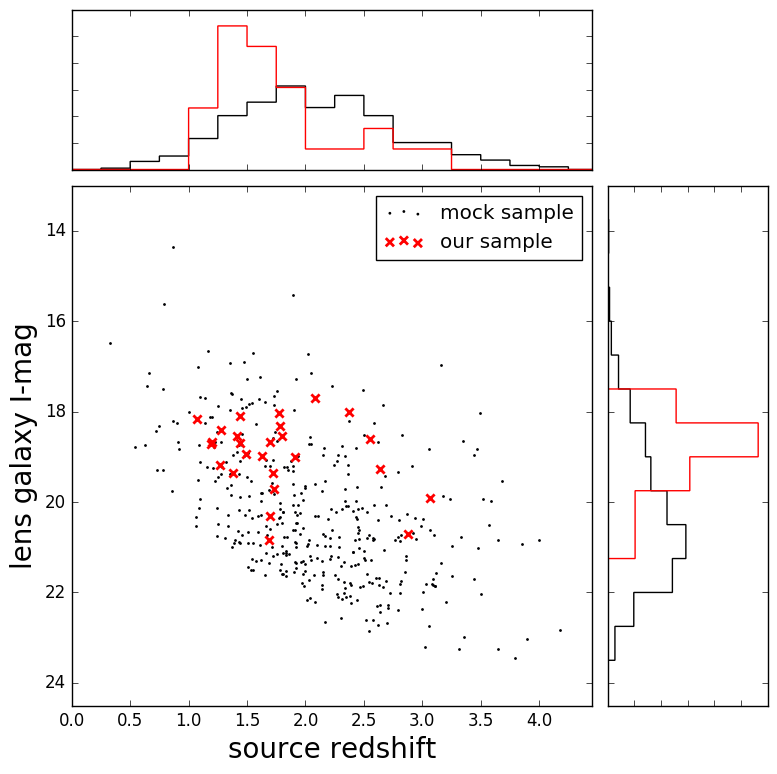}
    \caption{Plot of lensing galaxy $i$-band magnitude against source redshift for the 26 lensed quasars from this paper and paper II satisfying two \Gaia detections brighter than 20.7 for doubles, or three for quads, and image separations between 1 and 4 arcseconds. A complete mock sample with the same detection criteria is overlaid as black dots.}
    \label{fig:gaiasamplebias}
\end{figure}

We investigate the bias in our current sample of lenses discovered with \textit{Gaia}--including the lenses discovered in paper II--by comparing to a complete sample of mock lenses that would have images detected by \Gaia with image separations between 1 and 4 arcseconds (see Section \ref{predictions} for details of the mock lens sample). Figure \ref{fig:gaiasamplebias} shows the lensing galaxy brightness against source redshift for the 26 lenses confirmed in this series of papers that satisfy the above criteria. Also overlaid is the mock sample, demonstrating that we are clearly biased towards lensed quasars with bright lensing galaxies, as expected. We note that the nearly identical quasar pairs identified throughout our campaign might partly represent the missing systems with fainter lens galaxies. Though deeper imaging is required to reveal potential lensing galaxies in these systems, we are still limited by lack of spectroscopic follow-up time to confirm candidates without obvious lensing galaxies in the relatively shallow Pan-STARRS imaging. A bias towards lower-redshift sources is also present, explained both by the degeneracy with galaxy brightness and perhaps by prioritisation of candidates showing a $u$-band excess when such data are available. Figure \ref{fig:gaiasamplebias} also reveals that many lensed quasars detected by \textit{Gaia} are expected to have faint lensing galaxies that would not be uncovered given the depth of current Pan-STARRS imaging. These lenses could be targetted by modelling deeper data, such as Dark Energy Survey data, or by looking for quasar pairs with similar variability within \textit{Gaia} and/or LSST lightcurves. 

\begin{figure}
\includegraphics[width=0.5\textwidth]{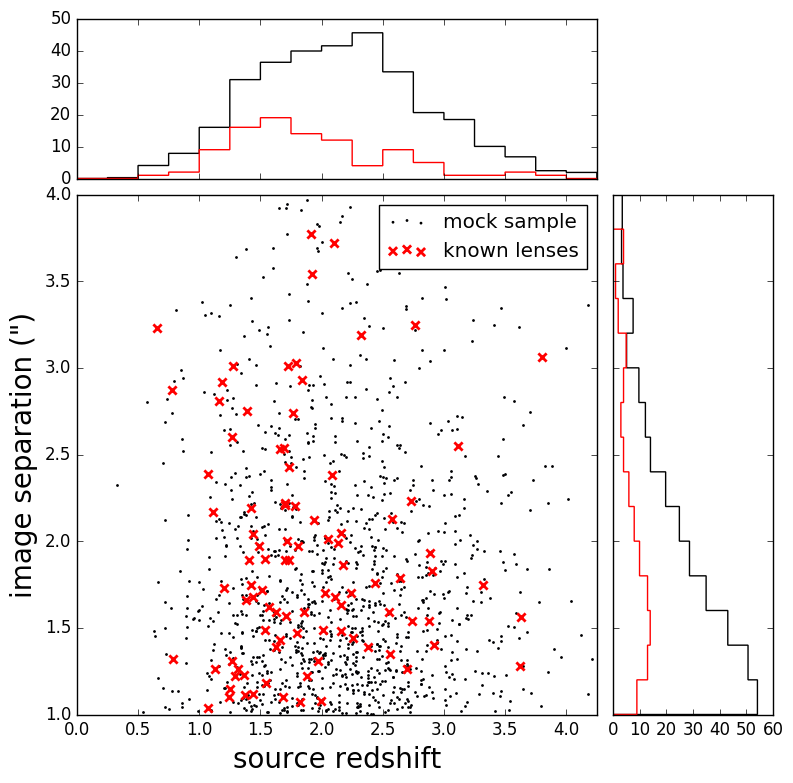}
    \caption{Plot of image separation against source redshift for all known lensed quasars satisfying |b|>20, image separations between 1 and 4 arcseconds, and 2 (at least 3) \Gaia detections brighter than 20.7 for doubles (quads). The image separations are in 0.2 arcsecond bins and the source redshifts in 0.25 bins. The mocks are plotted with a frequency 4 times that of the known lens area, since the OM10 catalogue is 100,000 square degrees and sky area outside of the galactic plane is 65.7\% of the full sky, i.e. 27100 square degrees. There are 96 known lenses that meet the stated criteria, while mocks predict 316 in total.}
    \label{fig:knownlensbias}
\end{figure}

Figure \ref{fig:knownlensbias} shows image separation versus source redshift for all known lensed quasars across the whole sky satisfying: 2 or at least 3 \Gaia detections for doubles and quads respectively, $G$<20.7 for each detection, |b|>20, and image separations between 1 and 4 arcseconds. This amounts to 96 such systems. We also plot the OM10 mock sample and histograms of the two parameters. The mock sample is normalised to the same area of sky in which the lenses are restricted to be. This shows that there are still $\sim$200 bright lensed quasars at modest separation left to be discovered, with $\sim$316 in total. These missing lenses are certainly being detected by \textit{Gaia} since it detects all components of known lenses and pairs of point sources at these magnitudes and separations. The majority of these undiscovered lenses will be in the Southern hemisphere, as imaging and spectroscopic surveys for this area of sky are recent or yet to begin.

\begin{figure}
\includegraphics[width=0.5\textwidth]{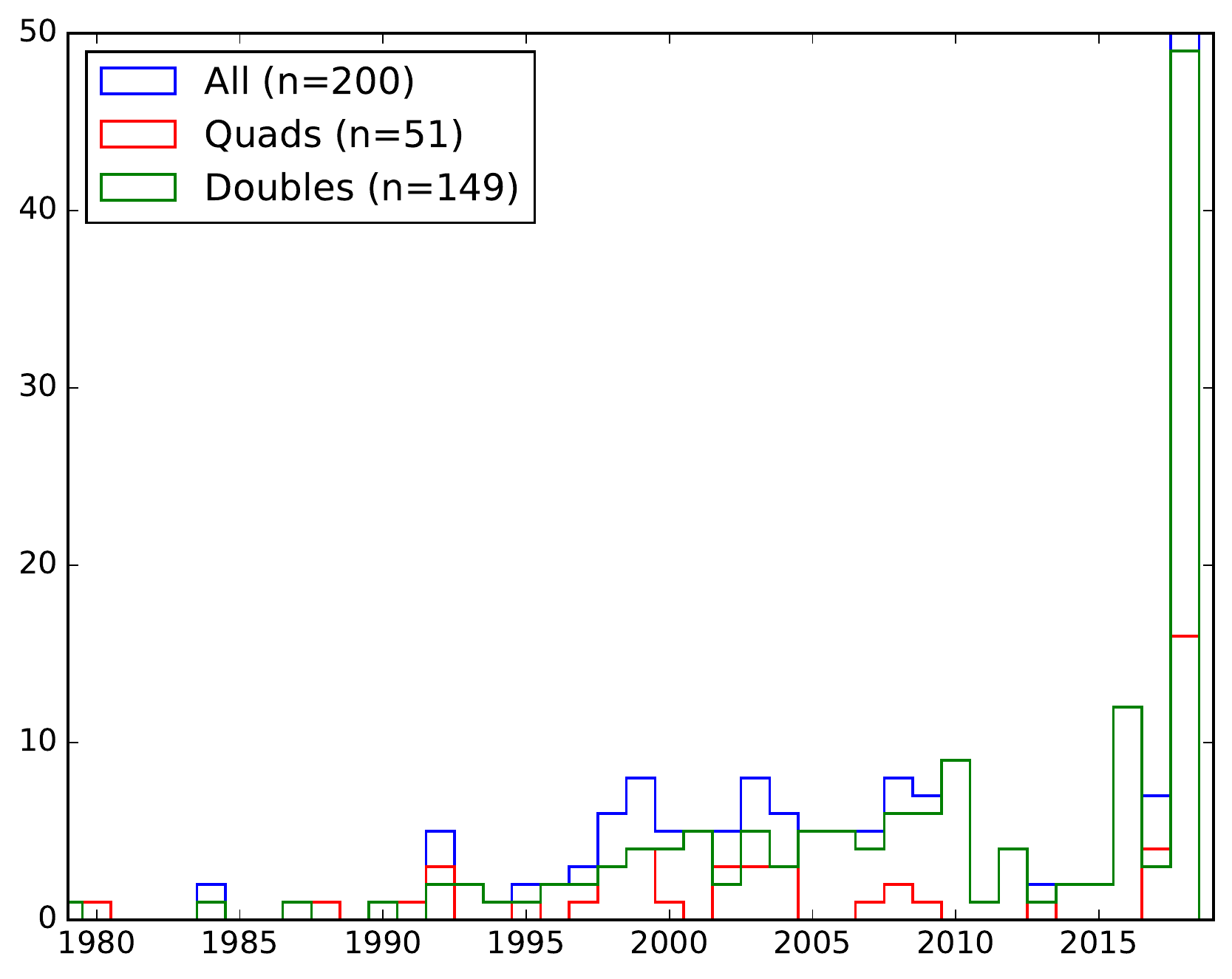}
    \caption{Number of published lensed quasars discovered each year, split into doubles and quads.}
    \label{fig:discoveryrate}
\end{figure}

To manage the rapidly increasing number of lensed quasars (Figure \ref{fig:discoveryrate}), we have created an online database that will be kept up-to-date with all known lensed quasars. Each lens has multi-band imaging cutouts, a variety of catalogue information, and relevant survey links.

\section{Conclusions} \label{conclusion}

We have presented the discovery of 22 new gravitationally lensed quasars with resolved spectra from ISIS on the WHT. Two lens candidates are very likely to be lensed quasars, however we have not been able to obtain spatially resolved spectra of the components. In total, we obtained a 70-90\% success rate in confirming gravitationally lensed quasars, while several candidates require deeper observations to understand their nature. The particularly low contaminant rate from a purely photometric selection arises from our ability to remove contaminants using \Gaia catalogue information and modelling of unWISE pixels. 

The lens sample presented here and \citet{lemon2018} is clearly a biased and incomplete sample, missing lensed quasars with fainter lensing galaxies and/or higher source redshifts. We have over 30 high-quality candidates left to follow-up that will begin to fill this missing parameter space. Future \Gaia data releases will help confirm these lenses efficiently via: more precise proper motion measurements, autocorrelation searches based on \Gaia lightcurves, and 2-D PSF reconstruction techniques to search for lensing galaxies directly in the \Gaia data \citep{harrison2011}. 

Building a complete sample of lensed quasars from \Gaia at modest image separation (>1 arcsecond) and from uncrowded fields (|b|>20 across the whole sky) will provide a useful statistical sample of around 350 lensed quasars, an order of magnitude greater than SQLS, which can be used to impose constraints on cosmological parameters and massive galaxy evolution \citep{oguri2012,finet2016}. We presently only know of approximately 100 lensed quasars in this sample, so an effort must be made to confirm the remaining systems.

\section*{Acknowledgements}
We would like to thank Elm\'e Breedt for her support with observations at the WHT, Paul Schechter for sharing his modelling of J1817+2729 and convincing us to observe the quad system, and Edi Rusu for helpful discussion regarding J1537-3010 and J1817+2729. We would also like to thank Patrick Hall for suggesting an updated redshift for J1418-1610. CAL and RGM would like to thank the STFC, and MWA acknowledges support from the STFC in the form of an Ernest Rutherford Fellowship. This work has made use of data from the European Space Agency (ESA) mission {\it Gaia} (\url{https://www.cosmos.esa.int/gaia}), processed by the {\it Gaia} Data Processing and Analysis Consortium (DPAC, \url{https://www.cosmos.esa.int/web/gaia/dpac/consortium}). Funding for the DPAC has been provided by national institutions, in particular the institutions participating in the {\it Gaia} Multilateral Agreement. This publication makes use of data products from the Wide-field Infrared Survey Explorer, which is a joint project of the University of California, Los Angeles, and the Jet Propulsion Laboratory/California Institute of Technology, funded by the National Aeronautics and Space Administration.





\bibliographystyle{mnras}
\bibliography{papers} 



\appendix
\begin{table*}
  \centering
  \caption{Pan-STARRS astrometry and photometry of all observed candidates. Magnitudes are in the AB sytem.}
  \label{tab:astrophotometry}
  \begin{tabular}{ccccccccc}
  \hline
  & component & $\alpha$ ($\arcsec$) & $\delta$ ($\arcsec$) & $g$ & $r$ & $i$ & $z$ & $Y$\\
  \hline
J0013+5119 & A & -0.83 $\pm$ 0.01 & 1.55 $\pm$ 0.02 & 20.86 $\pm$ 0.01 & 20.62 $\pm$ 0.01 & 20.49 $\pm$ 0.01 & 20.39 $\pm$ 0.02 & 20.17 $\pm$ 0.02\\
           & B & -0.02 $\pm$ 0.01 & -1.28 $\pm$ 0.02 & 21.31 $\pm$ 0.01 & 20.77 $\pm$ 0.02 & 20.54 $\pm$ 0.02 & 20.36 $\pm$ 0.03 & 20.29 $\pm$ 0.05\\
           & G & 0.00 $\pm$ 0.01 & 0.00 $\pm$ 0.02 & 20.55 $\pm$ 0.02 & 19.11 $\pm$ 0.02 & 18.41 $\pm$ 0.02 & 18.11 $\pm$ 0.02 & 17.81 $\pm$ 0.02\\
\hline
J0047+2514 & A & 0.77 $\pm$ 0.02 & -1.05 $\pm$ 0.02 & 21.03 $\pm$ 0.01 & 20.68 $\pm$ 0.01 & 20.74 $\pm$ 0.02 & 20.8 $\pm$ 0.03 & 20.75 $\pm$ 0.04\\
           & B & -0.21 $\pm$ 0.02 & 0.39 $\pm$ 0.02 & 21.36 $\pm$ 0.05 & 21.47 $\pm$ 0.12 & 21.74 $\pm$ 0.28 & --- & --- \\
           & G & 0.00 $\pm$ 0.02 & 0.00 $\pm$ 0.02 & 21.2 $\pm$ 0.07 & 19.34 $\pm$ 0.03 & 18.68 $\pm$ 0.03 & 18.44 $\pm$ 0.03 & 18.28 $\pm$ 0.05\\
\hline
J0102+2445 & A & 1.73 $\pm$ 0.01 & -0.71 $\pm$ 0.01 & 19.37 $\pm$ 0.01 & 19.35 $\pm$ 0.01 & 19.21 $\pm$ 0.01 & 19.0 $\pm$ 0.01 & 19.11 $\pm$ 0.01\\
           & B & -0.23 $\pm$ 0.01 & 0.34 $\pm$ 0.01 & 20.9 $\pm$ 0.02 & 20.74 $\pm$ 0.05 & 20.69 $\pm$ 0.06 & 20.04 $\pm$ 0.04 & 20.37 $\pm$ 0.08\\
           & G & 0.00 $\pm$ 0.01 & 0.00 $\pm$ 0.01 & 19.69 $\pm$ 0.02 & 18.45 $\pm$ 0.02 & 17.7 $\pm$ 0.02 & 17.59 $\pm$ 0.02 & 17.36 $\pm$ 0.02\\
\hline
J0124-0033 & A & -0.84 $\pm$ 0.02 & -1.06 $\pm$ 0.02 & 21.59 $\pm$ 0.02 & 21.15 $\pm$ 0.02 & 21.7 $\pm$ 0.03 & 21.93 $\pm$ 0.07 & 21.45 $\pm$ 0.1\\
           & B & 0.30 $\pm$ 0.02 & 0.56 $\pm$ 0.02 & 21.14 $\pm$ 0.06 & 21.21 $\pm$ 0.07 & 21.11 $\pm$ 0.06 & 21.03 $\pm$ 0.16 & 20.68 $\pm$ 0.19\\
           & G & 0.00 $\pm$ 0.01 & 0.00 $\pm$ 0.02 & 19.92 $\pm$ 0.03 & 18.64 $\pm$ 0.01 & 18.2 $\pm$ 0.01 & 17.91 $\pm$ 0.01 & 17.75 $\pm$ 0.02\\
\hline
J0203+1612 & A & 0.85 $\pm$ 0.02 & 1.53 $\pm$ 0.02 & 21.22 $\pm$ 0.01 & 20.53 $\pm$ 0.01 & 20.37 $\pm$ 0.01 & 19.93 $\pm$ 0.01 & 20.02 $\pm$ 0.02\\
           & B & -0.46 $\pm$ 0.02 & -0.87 $\pm$ 0.02 & 22.08 $\pm$ 0.03 & 21.11 $\pm$ 0.03 & 20.97 $\pm$ 0.04 & 20.67 $\pm$ 0.05 & 20.47 $\pm$ 0.1\\
           & G & 0.00 $\pm$ 0.02 & 0.00 $\pm$ 0.03 & 21.71 $\pm$ 0.04 & 20.22 $\pm$ 0.04 & 19.3 $\pm$ 0.04 & 18.81 $\pm$ 0.04 & 18.78 $\pm$ 0.05\\
\hline
J0228+3953 & A & 1.03 $\pm$ 0.01 & 0.19 $\pm$ 0.01 & 19.99 $\pm$ 0.02 & 19.76 $\pm$ 0.03 & 19.9 $\pm$ 0.03 & 19.32 $\pm$ 0.04 & 19.32 $\pm$ 0.04\\
           & B & -0.45 $\pm$ 0.01 & -0.31 $\pm$ 0.01 & 20.65 $\pm$ 0.02 & 20.36 $\pm$ 0.04 & 20.15 $\pm$ 0.04 & 19.92 $\pm$ 0.05 & 19.87 $\pm$ 0.09\\
           & G & 0.00 $\pm$ 0.03 & 0.00 $\pm$ 0.03 & --- & 21.05 $\pm$ 0.19 & 20.66 $\pm$ 0.17 & 19.85 $\pm$ 0.16 & 19.54 $\pm$ 0.18\\
\hline
J1238+2846 & A & -0.22 $\pm$ 0.01 & 1.54 $\pm$ 0.02 & 20.76 $\pm$ 0.01 & 20.62 $\pm$ 0.01 & 20.52 $\pm$ 0.01 & 20.49 $\pm$ 0.02 & 20.39 $\pm$ 0.02\\
           & B & 0.22 $\pm$ 0.01 & -0.84 $\pm$ 0.02 & 20.71 $\pm$ 0.01 & 20.64 $\pm$ 0.01 & 20.47 $\pm$ 0.01 & 20.5 $\pm$ 0.03 & 20.09 $\pm$ 0.04\\
           & G & 0.00 $\pm$ 0.05 & 0.00 $\pm$ 0.07 & --- & 22.08 $\pm$ 0.16 & 21.38 $\pm$ 0.15 & 20.28 $\pm$ 0.15 & 20.09 $\pm$ 0.12\\
\hline
J1307+0642 & A & 2.66 $\pm$ 0.01 & -1.17 $\pm$ 0.01 & 20.39 $\pm$ 0.01 & 20.16 $\pm$ 0.01 & 20.03 $\pm$ 0.01 & 19.83 $\pm$ 0.01 & 19.83 $\pm$ 0.01\\
           & B & -0.53 $\pm$ 0.01 & 0.29 $\pm$ 0.01 & 21.78 $\pm$ 0.05 & 21.24 $\pm$ 0.07 & 21.27 $\pm$ 0.09 & 20.88 $\pm$ 0.1 & 20.79 $\pm$ 0.17\\
           & G & 0.00 $\pm$ 0.01 & 0.00 $\pm$ 0.01 & 19.01 $\pm$ 0.01 & 17.58 $\pm$ 0.01 & 17.02 $\pm$ 0.01 & 16.75 $\pm$ 0.01 & 16.47 $\pm$ 0.01\\
\hline
J1418-1610 & A & 0.00 $\pm$ 0.01 & 0.00 $\pm$ 0.01 & 18.9 $\pm$ 0.01 & 18.57 $\pm$ 0.01 & 18.45 $\pm$ 0.01 & 18.64 $\pm$ 0.01 & 18.43 $\pm$ 0.01\\
           & B & 1.91 $\pm$ 0.01 & 1.47 $\pm$ 0.01 & 19.23 $\pm$ 0.0 & 18.92 $\pm$ 0.0 & 18.93 $\pm$ 0.0 & 18.88 $\pm$ 0.0 & 18.89 $\pm$ 0.0\\
\hline
J1428+0500 & A & 0.00 $\pm$ 0.01 & 0.00 $\pm$ 0.01 & 20.17 $\pm$ 0.02 & 19.91 $\pm$ 0.01 & 19.94 $\pm$ 0.02 & 19.86 $\pm$ 0.04 & 19.8 $\pm$ 0.09\\
           & B & 0.96 $\pm$ 0.01 & 2.01 $\pm$ 0.01 & 20.11 $\pm$ 0.02 & 20.03 $\pm$ 0.01 & 20.12 $\pm$ 0.02 & 19.96 $\pm$ 0.04 & 19.8 $\pm$ 0.08\\
\hline
J1515+3137 & A & 0.32 $\pm$ 0.01 & -0.90 $\pm$ 0.01 & 20.41 $\pm$ 0.01 & 20.23 $\pm$ 0.03 & 19.84 $\pm$ 0.02 & 19.67 $\pm$ 0.02 & 19.69 $\pm$ 0.04\\
           & B & -0.30 $\pm$ 0.01 & 0.49 $\pm$ 0.01 & 21.23 $\pm$ 0.05 & 21.09 $\pm$ 0.17 & 20.42 $\pm$ 0.13 & 20.27 $\pm$ 0.13 & 20.06 $\pm$ 0.18\\
           & G & 0.00 $\pm$ 0.03 & 0.00 $\pm$ 0.06 & 21.16 $\pm$ 0.07 & 19.87 $\pm$ 0.03 & 19.85 $\pm$ 0.06 & 19.46 $\pm$ 0.05 & 19.03 $\pm$ 0.06\\
\hline
J1518+4658 & A & -0.31 $\pm$ 0.01 & 0.78 $\pm$ 0.01 & 19.86 $\pm$ 0.01 & 19.67 $\pm$ 0.01 & 19.72 $\pm$ 0.02 & 19.46 $\pm$ 0.03 & 19.29 $\pm$ 0.04\\
           & B & -0.18 $\pm$ 0.01 & -0.57 $\pm$ 0.01 & 21.15 $\pm$ 0.04 & 20.97 $\pm$ 0.08 & 21.08 $\pm$ 0.12 & 20.39 $\pm$ 0.09 & 20.73 $\pm$ 0.17\\
           & G & 0.00 $\pm$ 0.02 & 0.00 $\pm$ 0.03 & 21.61 $\pm$ 0.13 & 20.25 $\pm$ 0.05 & 19.56 $\pm$ 0.03 & 19.57 $\pm$ 0.07 & 19.25 $\pm$ 0.07\\
\hline
J1524+4801 & A & -1.19 $\pm$ 0.02 & -0.91 $\pm$ 0.01 & 21.08 $\pm$ 0.01 & 20.62 $\pm$ 0.01 & 20.05 $\pm$ 0.01 & 19.96 $\pm$ 0.01 & 19.67 $\pm$ 0.01\\
           & B & 0.83 $\pm$ 0.02 & -0.00 $\pm$ 0.01 & 22.18 $\pm$ 0.07 & 21.77 $\pm$ 0.14 & 21.0 $\pm$ 0.06 & 20.77 $\pm$ 0.07 & 20.39 $\pm$ 0.07\\
           & G & 0.00 $\pm$ 0.02 & 0.00 $\pm$ 0.01 & 20.52 $\pm$ 0.03 & 19.26 $\pm$ 0.03 & 18.73 $\pm$ 0.03 & 18.53 $\pm$ 0.03 & 18.27 $\pm$ 0.07\\
\hline
J1537-3010 & A & -1.42 $\pm$ 0.01 & -0.82 $\pm$ 0.01 & 20.16 $\pm$ 0.01 & 20.08 $\pm$ 0.01 & 19.79 $\pm$ 0.01 & 19.8 $\pm$ 0.02 & 19.65 $\pm$ 0.02\\
           & B & 0.52 $\pm$ 0.01 & -1.07 $\pm$ 0.01 & 20.45 $\pm$ 0.02 & 20.38 $\pm$ 0.02 & 20.0 $\pm$ 0.03 & 20.02 $\pm$ 0.04 & 19.89 $\pm$ 0.05\\
           & C & 1.38 $\pm$ 0.01 & 0.90 $\pm$ 0.01 & 20.23 $\pm$ 0.01 & 20.24 $\pm$ 0.01 & 19.91 $\pm$ 0.01 & 19.89 $\pm$ 0.02 & 19.69 $\pm$ 0.02\\
           & D & -0.76 $\pm$ 0.01 & 1.02 $\pm$ 0.01 & 21.24 $\pm$ 0.03 & 20.99 $\pm$ 0.03 & 20.65 $\pm$ 0.04 & 20.48 $\pm$ 0.05 & 20.45 $\pm$ 0.06\\
           & G & 0.00 $\pm$ 0.04 & 0.00 $\pm$ 0.06 & 20.63 $\pm$ 0.07 & 20.18 $\pm$ 0.07 & 19.35 $\pm$ 0.06 & 18.98 $\pm$ 0.06 & 18.78 $\pm$ 0.05\\
\hline
J1553+3149 & A & -1.08 $\pm$ 0.01 & -0.38 $\pm$ 0.01 & 18.54 $\pm$ 0.01 & 18.47 $\pm$ 0.01 & 18.3 $\pm$ 0.02 & 18.14 $\pm$ 0.02 & 18.06 $\pm$ 0.04\\
           & B & 0.37 $\pm$ 0.01 & 0.30 $\pm$ 0.01 & 19.55 $\pm$ 0.02 & 19.65 $\pm$ 0.02 & 19.38 $\pm$ 0.02 & 19.27 $\pm$ 0.03 & 19.12 $\pm$ 0.05\\
           & G & 0.00 $\pm$ 0.02 & 0.00 $\pm$ 0.02 & 21.18 $\pm$ 0.26 & 19.2 $\pm$ 0.05 & 18.61 $\pm$ 0.05 & 18.39 $\pm$ 0.05 & 18.32 $\pm$ 0.07\\
\hline
J1554+5817 & A & 0.00 $\pm$ 0.01 & 0.00 $\pm$ 0.01 & 19.02 $\pm$ 0.01 & 19.16 $\pm$ 0.01 & 18.69 $\pm$ 0.01 & 18.86 $\pm$ 0.01 & 18.99 $\pm$ 0.01\\
           & B & -1.31 $\pm$ 0.01 & 0.47 $\pm$ 0.01 & 19.96 $\pm$ 0.01 & 19.84 $\pm$ 0.0 & 19.54 $\pm$ 0.0 & 19.63 $\pm$ 0.01 & 19.74 $\pm$ 0.01\\
\hline
J1612+3920 & A & 0.75 $\pm$ 0.01 & 1.74 $\pm$ 0.01 & 20.24 $\pm$ 0.01 & 20.09 $\pm$ 0.01 & 20.06 $\pm$ 0.01 & 19.95 $\pm$ 0.01 & 19.93 $\pm$ 0.01\\
           & B & -0.44 $\pm$ 0.01 & -0.72 $\pm$ 0.01 & 21.81 $\pm$ 0.04 & 21.89 $\pm$ 0.07 & 21.55 $\pm$ 0.06 & 21.41 $\pm$ 0.09 & ---\\
           & G & 0.00 $\pm$ 0.01 & 0.00 $\pm$ 0.01 & 19.26 $\pm$ 0.01 & 18.22 $\pm$ 0.01 & 17.87 $\pm$ 0.01 & 17.52 $\pm$ 0.01 & 17.36 $\pm$ 0.01\\
\hline
J1616+1415 & A & -0.28 $\pm$ 0.01 & -0.86 $\pm$ 0.01 & 20.1 $\pm$ 0.01 & 19.74 $\pm$ 0.01 & 19.6 $\pm$ 0.01 & 19.57 $\pm$ 0.02 & 19.45 $\pm$ 0.03\\
           & B & 0.35 $\pm$ 0.01 & 0.57 $\pm$ 0.01 & 20.59 $\pm$ 0.02 & 20.26 $\pm$ 0.01 & 20.07 $\pm$ 0.02 & 19.95 $\pm$ 0.04 & 19.86 $\pm$ 0.04\\
           & G & 0.00 $\pm$ 0.05 & 0.00 $\pm$ 0.08 & 21.16 $\pm$ 0.08 & 21.19 $\pm$ 0.06 & 20.7 $\pm$ 0.09 & 20.02 $\pm$ 0.07 & 19.53 $\pm$ 0.09\\

\hline
	\end{tabular}
\end{table*}
\addtocounter{table}{-1}
\begin{table*}
  \centering
  \caption{continued}
  \label{tab:astrophotometry}
  \begin{tabular}{ccccccccc}
  \hline
  & component & $\alpha$ ($\arcsec$) & $\delta$ ($\arcsec$) & $g$ & $r$ & $i$ & $z$ & $Y$\\
  \hline
\hline
J1623+7533 & A & -0.64 $\pm$ 0.01 & -1.07 $\pm$ 0.02 & --- & 19.81 $\pm$ 0.01 & 19.73 $\pm$ 0.01 & 19.57 $\pm$ 0.01 & ---\\
           & B & 0.03 $\pm$ 0.01 & 0.63 $\pm$ 0.02 & --- & 20.55 $\pm$ 0.03 & 20.68 $\pm$ 0.06 & 20.23 $\pm$ 0.06 & ---\\
           & G & 0.00 $\pm$ 0.02 & 0.00 $\pm$ 0.02 & --- & 19.87 $\pm$ 0.04 & 19.27 $\pm$ 0.04 & 19.11 $\pm$ 0.04 & ---\\
\hline
J1627-0224 & A & -0.86 $\pm$ 0.01 & 2.74 $\pm$ 0.01 & 19.72 $\pm$ 0.01 & 19.53 $\pm$ 0.01 & 19.13 $\pm$ 0.01 & 19.11 $\pm$ 0.01 & 19.02 $\pm$ 0.01\\
           & B & 0.75 $\pm$ 0.01 & -0.66 $\pm$ 0.01 & 20.51 $\pm$ 0.0 & 20.32 $\pm$ 0.01 & 19.93 $\pm$ 0.01 & 19.95 $\pm$ 0.01 & 19.86 $\pm$ 0.02\\
           & G & 0.00 $\pm$ 0.01 & 0.00 $\pm$ 0.01 & 21.48 $\pm$ 0.05 & 20.09 $\pm$ 0.03 & 19.01 $\pm$ 0.03 & 18.54 $\pm$ 0.03 & 18.3 $\pm$ 0.03\\
\hline
J1641+1002 & A & 0.00 $\pm$ 0.01 & 0.00 $\pm$ 0.01 & 19.16 $\pm$ 0.01 & 19.01 $\pm$ 0.01 & 18.72 $\pm$ 0.01 & 18.75 $\pm$ 0.01 & 18.72 $\pm$ 0.01\\
           & B & 0.93 $\pm$ 0.01 & 1.49 $\pm$ 0.01 & 21.18 $\pm$ 0.0 & 20.92 $\pm$ 0.0 & 20.74 $\pm$ 0.0 & 20.65 $\pm$ 0.01 & 20.84 $\pm$ 0.02\\
\hline
J1653+5155 & A & 0.39 $\pm$ 0.01 & -1.06 $\pm$ 0.01 & 20.06 $\pm$ 0.01 & 19.91 $\pm$ 0.01 & --- & 20.09 $\pm$ 0.01 & --- \\
           & B & -0.41 $\pm$ 0.01 & 0.25 $\pm$ 0.01 & 20.95 $\pm$ 0.01 & 20.64 $\pm$ 0.03 & --- & 19.87 $\pm$ 0.03 & --- \\
           & G & 0.00 $\pm$ 0.02 & 0.00 $\pm$ 0.01 & 21.46 $\pm$ 0.03 & 20.1 $\pm$ 0.02 & --- & 19.07 $\pm$ 0.02 & --- \\
\hline
J1724+0807 & A & -1.32 $\pm$ 0.01 & -0.73 $\pm$ 0.01 & 19.59 $\pm$ 0.01 & 18.78 $\pm$ 0.01 & 18.38 $\pm$ 0.01 & 18.26 $\pm$ 0.01 & 18.1 $\pm$ 0.01\\
           & B & 0.73 $\pm$ 0.01 & -0.25 $\pm$ 0.01 & 21.0 $\pm$ 0.01 & 20.23 $\pm$ 0.02 & 19.86 $\pm$ 0.02 & 19.83 $\pm$ 0.03 & 19.66 $\pm$ 0.03\\
           & G & 0.00 $\pm$ 0.01 & 0.00 $\pm$ 0.01 & 20.97 $\pm$ 0.02 & 19.62 $\pm$ 0.01 & 19.03 $\pm$ 0.01 & 18.8 $\pm$ 0.02 & 18.62 $\pm$ 0.01\\
\hline
J1817+2729 & A & 0.67 $\pm$ 0.01 & 0.42 $\pm$ 0.01 & 19.26 $\pm$ 0.01 & 18.87 $\pm$ 0.01 & 18.5 $\pm$ 0.01 & 18.48 $\pm$ 0.01 & 18.24 $\pm$ 0.02\\
           & B & 0.83 $\pm$ 0.01 & -0.43 $\pm$ 0.01 & 22.24 $\pm$ 0.12 & 21.81 $\pm$ 0.11 & 21.62 $\pm$ 0.12 & 21.7 $\pm$ 0.32 & 21.04 $\pm$ 0.23\\
           & C & -0.61 $\pm$ 0.01 & -0.84 $\pm$ 0.01 & 20.34 $\pm$ 0.01 & 19.93 $\pm$ 0.01 & 19.61 $\pm$ 0.01 & 19.59 $\pm$ 0.01 & 19.31 $\pm$ 0.02\\
           & D & -0.61 $\pm$ 0.01 & 0.63 $\pm$ 0.01 & 21.5 $\pm$ 0.04 & 20.81 $\pm$ 0.03 & 20.47 $\pm$ 0.04 & 20.29 $\pm$ 0.05 & 19.83 $\pm$ 0.04\\
           & G & 0.00 $\pm$ 0.03 & 0.00 $\pm$ 0.03 & 22.73 $\pm$ 0.28 & 20.63 $\pm$ 0.07 & 19.92 $\pm$ 0.05 & 19.33 $\pm$ 0.06 & 19.23 $\pm$ 0.08\\
\hline
J1949+7732 & A & 0.51 $\pm$ 0.01 & -0.51 $\pm$ 0.01 & 18.96 $\pm$ 0.03 & 18.56 $\pm$ 0.03 & 18.82 $\pm$ 0.04 & 18.7 $\pm$ 0.05 & 18.73 $\pm$ 0.1\\
           & B & -0.77 $\pm$ 0.01 & 0.44 $\pm$ 0.01 & 19.88 $\pm$ 0.02 & 19.69 $\pm$ 0.03 & 19.51 $\pm$ 0.04 & 19.31 $\pm$ 0.05 & 19.33 $\pm$ 0.09\\
           & G & 0.00 $\pm$ 0.02 & 0.00 $\pm$ 0.02 & --- & 19.55 $\pm$ 0.1 & 18.98 $\pm$ 0.1 & 18.59 $\pm$ 0.09 & 18.32 $\pm$ 0.11\\
\hline
J2014-3024 & A & 0.00 $\pm$ 0.01 & 0.00 $\pm$ 0.01 & 19.14 $\pm$ 0.01 & 18.95 $\pm$ 0.01 & 18.72 $\pm$ 0.01 & 18.56 $\pm$ 0.04 & 18.23 $\pm$ 0.07\\
           & B & 1.00 $\pm$ 0.01 & 0.17 $\pm$ 0.01 & 19.49 $\pm$ 0.01 & 19.17 $\pm$ 0.01 & 19.03 $\pm$ 0.01 & 18.91 $\pm$ 0.04 & 18.91 $\pm$ 0.06\\
           & C & 2.18 $\pm$ 0.01 & 1.18 $\pm$ 0.01 & 19.36 $\pm$ 0.01 & 19.02 $\pm$ 0.01 & 18.86 $\pm$ 0.0 & 18.82 $\pm$ 0.03 & 18.54 $\pm$ 0.06\\
\hline
J2032-2358 & A & 0.00 $\pm$ 0.01 & 0.00 $\pm$ 0.01 & 19.48 $\pm$ 0.01 & 19.21 $\pm$ 0.01 & 18.86 $\pm$ 0.01 & 18.71 $\pm$ 0.03 & 18.84 $\pm$ 0.06\\
           & B & 1.91 $\pm$ 0.01 & 0.09 $\pm$ 0.01 & 19.8 $\pm$ 0.01 & 19.56 $\pm$ 0.01 & 19.25 $\pm$ 0.01 & 19.33 $\pm$ 0.02 & 19.44 $\pm$ 0.05\\
\hline
J2132+2603 & A & 0.83 $\pm$ 0.01 & 0.76 $\pm$ 0.01 & 19.43 $\pm$ 0.03 & 19.48 $\pm$ 0.02 & 19.44 $\pm$ 0.03 & 19.31 $\pm$ 0.04 & 18.87 $\pm$ 0.08\\
           & B & -0.51 $\pm$ 0.01 & -0.39 $\pm$ 0.01 & 20.63 $\pm$ 0.05 & 20.74 $\pm$ 0.04 & 20.61 $\pm$ 0.07 & 20.44 $\pm$ 0.09 & 20.01 $\pm$ 0.14\\
           & G & 0.00 $\pm$ 0.04 & 0.00 $\pm$ 0.04 & 21.3 $\pm$ 0.24 & 20.77 $\pm$ 0.13 & 19.75 $\pm$ 0.08 & 19.37 $\pm$ 0.08 & 19.01 $\pm$ 0.12\\
\hline
J2145+6345 & A & 0.00 $\pm$ 0.01 & 0.00 $\pm$ 0.01 & 17.79 $\pm$ 0.01 & 17.28 $\pm$ 0.01 & 16.71 $\pm$ 0.01 & 16.69 $\pm$ 0.01 & 16.54 $\pm$ 0.01\\
           & B & 0.33 $\pm$ 0.01 & -0.56 $\pm$ 0.01 & 17.7 $\pm$ 0.01 & 17.04 $\pm$ 0.0 & 16.51 $\pm$ 0.0 & 16.46 $\pm$ 0.0 & 16.26 $\pm$ 0.01\\
           & C & 1.87 $\pm$ 0.01 & -0.88 $\pm$ 0.01 & 18.82 $\pm$ 0.0 & 18.32 $\pm$ 0.0 & 17.82 $\pm$ 0.0 & 17.79 $\pm$ 0.0 & 17.52 $\pm$ 0.0\\
           & D & 1.37 $\pm$ 0.01 & 0.81 $\pm$ 0.01 & 19.14 $\pm$ 0.0 & 18.55 $\pm$ 0.0 & 18.08 $\pm$ 0.0 & 17.91 $\pm$ 0.0 & 17.72 $\pm$ 0.0\\
\hline
J2212+3144 & A & 0.00 $\pm$ 0.01 & 0.00 $\pm$ 0.01 & 19.6 $\pm$ 0.02 & 19.55 $\pm$ 0.01 & 19.17 $\pm$ 0.02 & 18.91 $\pm$ 0.03 & 19.18 $\pm$ 0.11\\
           & B & -0.69 $\pm$ 0.01 & 2.56 $\pm$ 0.01 & 20.21 $\pm$ 0.01 & 20.19 $\pm$ 0.01 & 19.79 $\pm$ 0.01 & 19.82 $\pm$ 0.02 & 20.02 $\pm$ 0.09\\
\hline
J2250+2117 & A & -1.48 $\pm$ 0.01 & -0.16 $\pm$ 0.01 & 18.58 $\pm$ 0.01 & 18.32 $\pm$ 0.01 & 18.02 $\pm$ 0.01 & 18.13 $\pm$ 0.02 & 18.12 $\pm$ 0.05\\
           & B & 0.37 $\pm$ 0.01 & 0.04 $\pm$ 0.01 & 20.39 $\pm$ 0.02 & 19.87 $\pm$ 0.02 & 19.72 $\pm$ 0.04 & 19.96 $\pm$ 0.07 & 19.72 $\pm$ 0.11\\
           & G & 0.00 $\pm$ 0.05 & 0.00 $\pm$ 0.02 & --- & --- & 19.71 $\pm$ 0.09 & 19.07 $\pm$ 0.09 & 19.32 $\pm$ 0.18\\
\hline
J2316+0610 & A & -1.41 $\pm$ 0.01 & 1.38 $\pm$ 0.01 & 20.35 $\pm$ 0.01 & 20.15 $\pm$ 0.01 & 19.95 $\pm$ 0.01 & 19.71 $\pm$ 0.01 & 19.96 $\pm$ 0.01\\
           & B & 0.37 $\pm$ 0.01 & -0.06 $\pm$ 0.01 & 20.99 $\pm$ 0.02 & 20.7 $\pm$ 0.04 & 20.74 $\pm$ 0.09 & 20.32 $\pm$ 0.08 & 20.55 $\pm$ 0.2\\
           & G & 0.00 $\pm$ 0.02 & 0.00 $\pm$ 0.02 & --- & 20.93 $\pm$ 0.07 & 19.81 $\pm$ 0.05 & 19.55 $\pm$ 0.05 & 19.4 $\pm$ 0.1\\
\hline
J2350+3654 & A & -0.68 $\pm$ 0.02 & -2.00 $\pm$ 0.02 & 21.17 $\pm$ 0.01 & 21.03 $\pm$ 0.01 & 20.82 $\pm$ 0.01 & 20.72 $\pm$ 0.01 & 20.77 $\pm$ 0.02\\
           & B & 1.14 $\pm$ 0.02 & 0.77 $\pm$ 0.02 & 21.4 $\pm$ 0.01 & 21.52 $\pm$ 0.04 & 21.1 $\pm$ 0.02 & 20.76 $\pm$ 0.03 & 21.19 $\pm$ 0.08\\
           & C & -1.58 $\pm$ 0.02 & 0.44 $\pm$ 0.02 & 22.74 $\pm$ 0.04 & 22.8 $\pm$ 0.12 & 21.9 $\pm$ 0.04 & 21.79 $\pm$ 0.06 & 21.16 $\pm$ 0.06\\
           & G & 0.00 $\pm$ 0.02 & 0.00 $\pm$ 0.02 & 20.38 $\pm$ 0.01 & 18.96 $\pm$ 0.01 & 18.06 $\pm$ 0.01 & 17.72 $\pm$ 0.01 & 17.55 $\pm$ 0.01\\

\hline
	\end{tabular}
\end{table*}


\bsp	
\label{lastpage}
\end{document}